\documentclass[aps,prb,twocolumn,english,showpacs,showkeys,floatfix]{revtex4-1}

\usepackage{dcolumn}
\usepackage{bm}

\usepackage{graphicx} 
\usepackage{epsfig}
\usepackage{amsfonts}
\usepackage[figures right]{rotating}
\usepackage{amssymb}
\usepackage{amsmath}
\usepackage{amsmath}
\usepackage{graphicx}
\usepackage{hyperref}
\usepackage{color}
\usepackage{soul}
\usepackage{tabularx}
\usepackage{array}
\usepackage{epstopdf}  
\usepackage{verbatim}

\begin{document}

\title{Mixing of $t_{2g}$-$e_g$ orbitals in 4d and 5d transition metal oxides}

\author{Georgios L. Stamokostas}
\email{geostam21@gmail.com}
\affiliation{Department of Physics, The University of Texas at Austin, Austin, TX, 78712, USA}
\author{Gregory A. Fiete}
\affiliation{Department of Physics, The University of Texas at Austin, Austin, TX, 78712, USA}

\date{\today}

\begin{abstract}
Using exact diagonalization, we study the spin-orbit coupling and interaction-induced mixing between $t_{2g}$ and $e_g$ $d$-orbital states in a cubic crystalline environment, as commonly occurs in transition metal oxides.  We make a direct comparison with the widely used $t_{2g}$ only or  $e_g$ only model, depending on electronic filling. We consider all electron fillings of the $d$-shell and compute the total magnetic moment, the spin, the occupancy of each orbital, and the effective spin-orbit coupling strength (renormalized through interaction effects) in terms of the bare interaction parameters, spin-orbit coupling, and crystal field splitting, focusing on the parameter ranges relevant to 4d and 5d transition metal oxides.  In various limits we provide perturbative results consistent with our numerical calculations.  We find that the $t_{2g}$-$e_g$ mixing can be large, with up to 20\% occupation of orbitals that are nominally ``empty", which has experimental implications for the interpretation of the branching ratio in experiments, and can impact the effective local moment Hamiltonian used to study magnetic phases and magnetic excitations in transition metal oxides.  Our results can aid the theoretical interpretation of experiments on these materials, which often fall in a regime of intermediate coupling with respect to electron-electron interactions.
\end{abstract}


\maketitle


\section{\label{sec:intro2}Introduction}
Transition metal oxides have undergone intensive study because of their remarkably rich phase diagrams and sensitivity to external fields, strain, disorder, and doping.\cite{Bednorz:rmp88,Lee:rmp06,Tokura:sci00,Gardner:rmp10}  High-temperature superconductors (e.g., cuprates) and colossal magnetoresistance materials (e.g., manganites) are two notable examples, but both of these have light transition elements drawn from the 3d series.\cite{TMO_book,Imada:rmp98}   On the other hand, the study of topological insulators in recent years\cite{Qi:rmp11,Hasan:rmp10,Moore:nat10,Ando:jpsj13}  has brought attention to the importance of large spin-orbit coupling, which may induce topological phase transitions in materials.  As a result, some focus has shifted to the heavier transition metals from the 4d and 5d series, which have significantly enhanced spin-orbit coupling relative to those in the 3d series.\cite{Krempa:arcm14,Rau:arcm16,Schaffer:rpp16}  

Iridates, in particular, have undergone much theoretical and experimental study.\cite{Krempa:arcm14,Rau:arcm16,Schaffer:rpp16}  An interesting body of theoretical studies has suggested that novel interaction-driven topological states in which the quantum numbers of the electron are fractionalized may appear.\cite{Maciejko:np15,Stern:arcmp16}  However, in some of the iridates even the nature of the conventional order, such as the magnetic order (and the underlying microscopic spin Hamiltonian), is not easy to determine,\cite{Chaloupka:prl10,Jiang:prb11,Kimchi:prb15,Takayama:prl15,Alpichshev:prl15,Williams:prb16,Biffin:prb14,Chern:prb17,Sizyuk:prb14}
in part due to the large neutron absorption cross-section which makes neutron scattering experiments challenging.\cite{Choi:prl12}  An experimental tool known as resonant inelastic X-ray scattering (RIXS) is particularly well suited to studies of the iridates.\cite{Ament:rmp11,Kotani:rmp01,Gretarsson:prl16,Lu:prl17,Moretti:prb15,Kim_Jungho:prl12,Moretti:prl14}  While there is some understanding of the microscopic details revealed in the RIXS signal, the theory is still under development.\cite{Savary:14}  Our work will facilitate that development.

A further challenge to understanding the iridates and other 4d/5d transition metal oxides is that the materials fall into a regime of comparable energy scales where it is difficult to argue {\em a priori} that a particular term in the Hamiltonian is small compared to the others: The typical kinetic energy, interaction energies, Hund's coupling, spin-orbit coupling, and crystal field splitting are all on the scale of an electron volt.\cite{Krempa:arcm14,Rau:arcm16,Schaffer:rpp16}  With respect to theoretical analysis, this means it is not clear if one should approach the iridates from a weak-coupling band-like description in which correlations are included within the band description,\cite{Hu:sr15,Chen_DMFT:prb15,Zhang_Huale:prl13,Zhang:prl17,Wan:prb11} or from the strong-coupling limit in which a local moment model\cite{Hozoi:prb14,Katukuri:prb12,Mohapatra:prb17,Kim_Jungho_2:prl12,Perkins:prb14,Svoboda:prb17,Meetei:prb15,Yuan:prb17,Laurell:prl17} is natural to describe the various types of magnetic orders that typically occur in the 4d/5d transition metal oxides (characteristic magnetic transition temperatures are on the order of 100K).\cite{Krempa:arcm14,Rau:arcm16,Schaffer:rpp16}  In this work, we start from an atomic limit of the transition metal ions and treat the interaction effects non-pertubatively using exact diagonalization.  In this way, we are able to work within an intermediate regime that reduces to a tight-binding-type Hamiltonian (for multiple ions) in the limit of vanishing interactions and a local moment model in the limit of strong interactions.

In a large class of transition metal oxides, the local oxygen environment of the transition metal ions is an octahedral cage (see Fig.~\ref{fig:octahedral_cages}) that produces a cubic environment that splits the $d$-orbitals into a lower lying triply degenerate $t_{2g}$ set of orbitals and a higher lying doubly-degenerate $e_g$ set of orbitals.  A feature that is shared by nearly all weak (aside from {\em ab initio} studies) {\em and} strong-coupling theoretical studies of the heavy transition metal oxides is that they assume the $t_{2g}$-$e_g$ mixing is negligible.\cite{Hozoi:prb14,Katukuri:prb12,Mohapatra:prb17,Kim_Jungho_2:prl12,Perkins:prb14,Svoboda:prb17,Meetei:prb15,Yuan:prb17,Laurell:prl17} In addition, many theoretical studies motivated by the iridates assume the infinite spin-orbit coupling limit which splits the $t_{2g}$ orbitals into a total angular moment $J_{\rm eff}=3/2$ and $J_{\rm eff}=1/2$ set of states (that do not mix).  For iridates with a nominal $d$-shell filling of 5 electrons, this results in a half-filled $J_{\rm eff}=1/2$ band, and thus reduces the Hamiltonian to a one-band model that often helps theoretical studies that rely on methods developed in the context of the cuprates.

In this work, we revisit the assumption of negligible $t_{2g}$-$e_g$ mixing and study the single ion limit in detail using exact diagonalization that allows a non-perturbative treatment of interaction effects.  We consider all $d$-shell fillings and find the neglect of  $t_{2g}$-$e_g$ mixing is not in general justified, with the greatest mixing occurring for fillings of 5,6, and 7 electrons.  Our work has implications for the interpretation of RIXS and X-ray absorption spectroscopy (XAS) data for the heavier elements with strong spin-orbit coupling, and the spectra of transition metal ions in oxides more generally. Our work can also be used as a more realistic starting point for determining the best form of the magnetic interactions between two nearby ions: Exchange interactions, exchange anistropies, and the size of local moments differ as a consequence of $t_{2g}$-$e_g$ mixing.

Our paper is organized as follows.  In Sec.~\ref{sec:crystal_field} we summarize the effects of a local cubic crystal field on the $d$-orbital level structure of a transition metal ion.  In Sec.~\ref{sec:T-P_and_beyond} we provide the details of the Hamiltonian with and without $t_{2g}$-$e_g$ mixing in the presence of spin-orbit coupling. In Sec.~\ref{sec:add_e-e_int}  and Sec.~\ref{sec:Model_and_calculations} we describe the interaction terms and conserved quantities of the full system we study, and in Sec.~\ref{sec:Results} we present the results of our exact diagonalization studies for all electron fillings.  We present the main conclusions of the work in Sec.~\ref{sec:Summary2}.

\section{\label{sec:crystal_field}Octahedral crystal fields}

\begin{figure}[h]
    \includegraphics[scale=0.5]{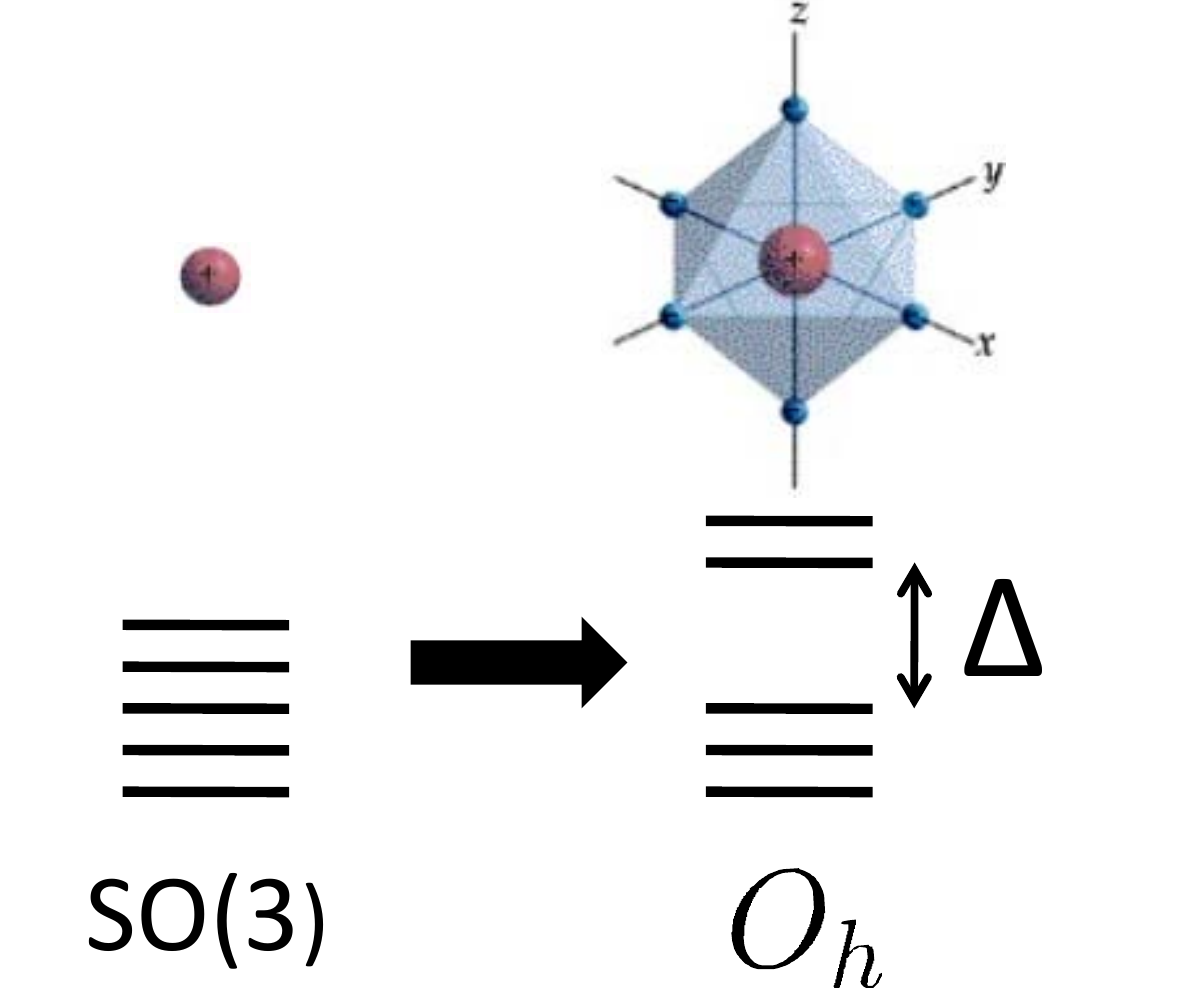}
    \caption{(Color online) Symmetry lowering and level splitting in a cubic crystal field environment.   A transition metal ion in free space has a full rotational SO(3) symmetry reduced to octahedral symmetry $O_h$.  The five-fold degenerate $d$-levels in the vacuum split into a lower-lying triply degenerate $t_{2g}$, and a higher-lying doubly degenerate $e_g$ set of levels, with an energy difference $\Delta$ (called the crystal field splitting) between them.}
    \label{fig:octahedral_cages}
 \end{figure}

\begin{figure}[h]
    \includegraphics[scale=0.5]{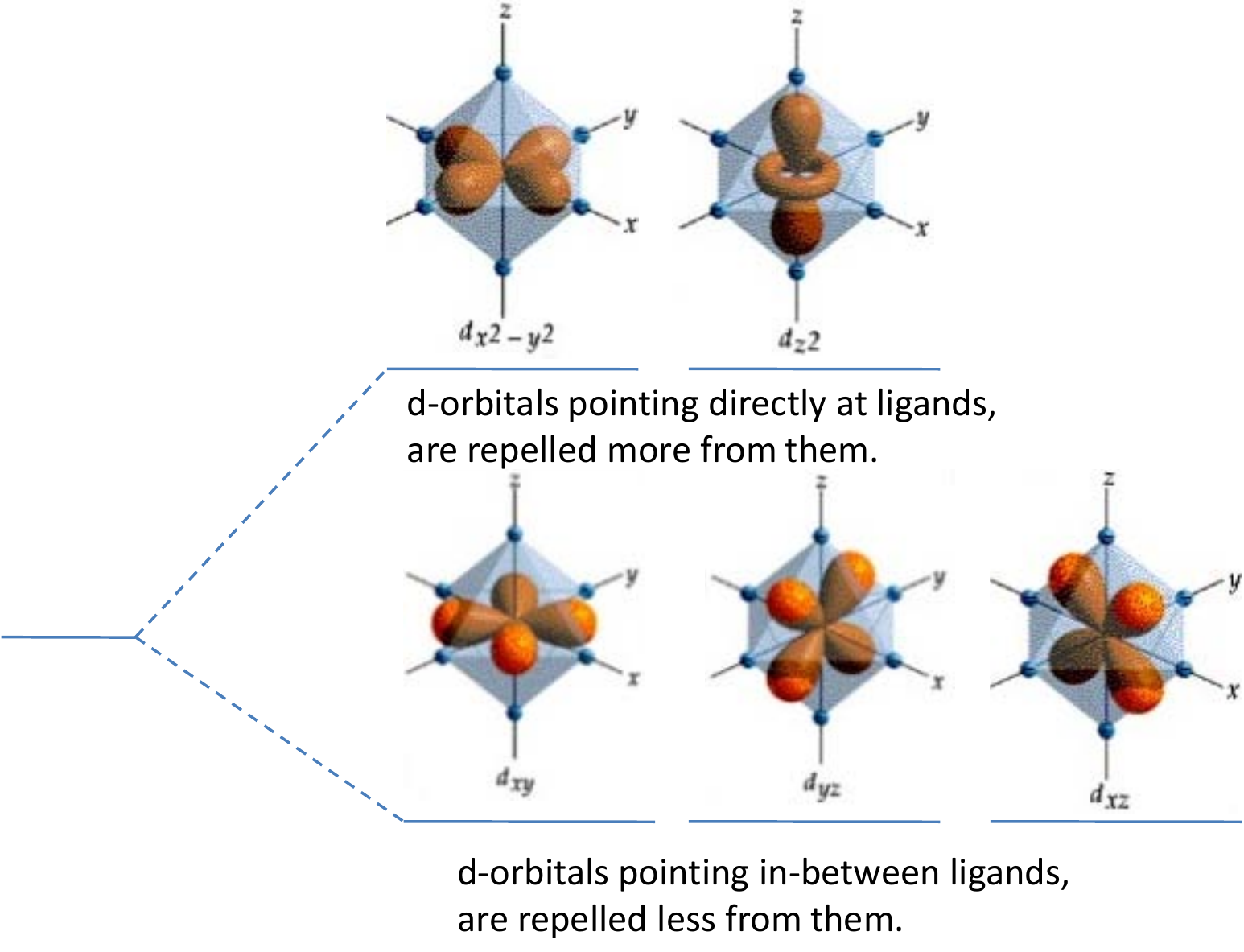}
    \caption{(Color online) The $t_{2g}$ wavefunctions have electron clouds pointing in between the point charges of the ligands, thus they repel less and have lower energy, compared to the $e_g$ states which point towards the oxygen ligands.}
    \label{fig:t2g_eg}
 \end{figure}

A transition metal ion in free space has rotational symmetry SO(3) and therefore five-fold degenerate $d$-orbitals. Frequently, transition metal ions in crystals are held inside regular octahedral cages, surrounded by ligands. A common type of these ligands is oxygen, which form the large class of transition metal oxides. When a free ion is placed inside an octahedral cage,  the symmetry is reduced from the full rotational SO(3) symmetry of the $d$-orbital states in the free space, to the symmetry group of the octahedron, SO(3)$\rightarrow O_h$. This consists of all the rotations which take the octahedron into itself.  Thus, $O_h$ is a subgroup of the rotation group: $O_h \subset $SO(3).  Hence, any representation of SO(3) provides a representation of $O_h$. However, irreducible representations of SO(3) will become reducible representations of $O_h$. Thus, the fivefold degeneracy of the $d$-states is lifted by the crystal field and the $d$-levels are split into a higher-lying two-fold degenerate $e_g$ and a lower-lying three-fold degenerate $t_{2g}$ manifold, as seen in Fig.\ref{fig:octahedral_cages}, where $\Delta$ is the energy difference between them. The oxygen ligands are approximated as point charges siting in the corners of the octahedral cages. The $t_{2g}$ $d$-orbital charge distributions point in between the point charges of the oxygens, and the $e_g$ states point towards the point charges, raising their energy relative to the $t_{2g}$ levels, as shown in Fig. \ref{fig:t2g_eg}.

 The $t_{2g}$ and $e_g$ orbitals are formed by linear combinations\cite{TMO_book} of the spherical harmonics $Y_l^m$, with the orbital angular momentum $l=2$.  The magnetic quantum number $m$ takes values from $-l$ to $l$. For $t_{2g}$ these orbitals states are:
\begin{equation}
\label{eq:t2g}
\begin{split}
d_{yz}&=-\frac{1}{i \sqrt{2}} (Y_2^1+Y_2^{-1}),\\
d_{zx}&=-\frac{1}{i \sqrt{2}} ( Y_2^1-Y_2^{-1}),\\
d_{xy}&=\frac{1}{i \sqrt{2}} (Y_2^2- Y_2^{-2}),
\end{split}
\end{equation}
and for $e_{g}$ they are: 
\begin{equation}
\label{eq:eg}
\begin{split}
d_{3z^2-r^2}&= Y_2^0,\\
d_{x^2-y^2}&=\frac{1}{ \sqrt{2}} (Y_2^2+Y_2^{-2}).
\end{split}
\end{equation}
 
The crystal field term in the Hamiltonian, $H_{\rm CF}$, can be written in a diagonal form as (taking the energy of the $t_{2g}$ states as the zero of energy),
\begin{equation}\label{eq:Hcf}
\begin{split}
H_{\rm CF}&=\sum_{\sigma=\pm 1/2}\Delta (\vert 3z^2-r^2,\sigma \rangle \langle 3z^2-r^2,\sigma \vert\\
&+\vert x^2-y^2,\sigma \rangle \langle x^2-y^2,\sigma \vert),
\end{split}
\end{equation}
where $\sigma=\pm1/2$ refers to the spin of the electron in a given orbital state.
 
 \section{\label{sec:T-P_and_beyond} Spin-orbit coupling in a crystal field}
The spin-orbit coupling strength is comparable to other energy scales in heavy transition metal oxides.\cite{Krempa:arcm14,Rau:arcm16,Schaffer:rpp16} In its presence the orbital angular momentum and spin angular momentum are no longer independently conserved quantities.  Moreover, the spin-orbit coupling can also induce mixing between the $t_{2g}$ and $e_g$ manifolds.

The matrix elements of orbital angular momentum $l$ for a single electron in the basis of the $t_{2g}$, Eq. \eqref{eq:t2g}, and $e_g$, Eq.\eqref{eq:eg}, states: $\{d_{yz},d_{zx},d_{xy},d_{3z^2-r^2},d_{x^2-y^2}\}$, and that of a
single electron in atomic $p$-orbitals in the basis $\{p_x,p_y,p_z\}$ are:\cite{sugano} 
\begin{equation}
\label{eq:lxmatrixelements}
 l_x=
 \begin{bmatrix}
\begin{array}{c c c|c c}
  0 & 0 & 0  & -\sqrt{3}i & -i \\ 
 0 & 0 & i  &0 &0\\
 0 & -i & 0 &0 &0 \\
 \hline
 \sqrt{3} i & 0 & 0 &0 &0\\
  i & 0 & 0 &0 &0
\end{array}
\end{bmatrix},
l_x^\prime=\begin{pmatrix}
\begin{array}{c c c}
 0 & 0 & 0   \\ 
 0 & 0 &- i  \\
 0 & i & 0 
\end{array}
\end{pmatrix},
\end{equation}
\begin{equation}
\label{eq:lymatrixelements}
 l_y=
\begin{bmatrix}
\begin{array}{c c c|c c}
 0 & 0 & -i  & 0 & 0 \\ 
 0 & 0 & 0  &\sqrt{3}i &-i\\
 i & 0 & 0 &0 &0 \\
 \hline
0 & -\sqrt{3} i & 0 &0 &0\\
  0 &  i & 0 &0 &0
\end{array}
\end{bmatrix},
l_y^\prime=\begin{pmatrix}
\begin{array}{c c c}
 0 & 0 & i   \\ 
 0 & 0 &0  \\
 -i & 0 & 0 
\end{array}
\end{pmatrix},
\end{equation}
\begin{equation}
\label{eq:lzmatrixelements}
 l_z=
\begin{bmatrix}
\begin{array}{c c c|c c}
 0 & i & 0  & 0 & 0\\ 
 -i & 0 & 0  &0 &0\\
 0 & 0  & 0 &0 & 2i \\
 \hline
 0 & 0 & 0 &0 &0\\
  0 & 0 & -2i &0 &0
\end{array}
\end{bmatrix},
l_z^\prime=\begin{pmatrix}
\begin{array}{c c c}
 0 & -i & 0   \\ 
 i & 0 &0  \\
 0 & 0 & 0 
\end{array}
\end{pmatrix}.
\end{equation}

By comparing the matrix elements of $l$ in the $t_{2g}$ states with those in the $p$-states in free atoms, one can map the former
$l$ = 2 $t_{2g}$-states onto the latter $p$-states with $l$ = 1 using the relation:
\begin{equation}
\label{eq:tpequiv}
\bm{l}(t_{2g})=-\bm{l}(p).
\end{equation}
This relation is called the T-P equivalence, \cite{sugano,fazekas} according to which the orbital angular momentum
in $t_{2g}$ states is partially quenched from $l=2$ to $l=1$. When the cubic crystal field
splitting is large, one can neglect the off-diagonal elements between $t_{2g}$ and $e_g$ manifolds and
the T-P equivalence can be conveniently used.  Note, however, that the spin-orbit coupling generally mixes the $t_{2g}$ and $e_g$ states so if the spin-orbit coupling is large enough compared to the crystal field splitting (and we will see it can be enhanced by electron-electron interactions) then the mixing may have non-negligible effects.

Using the expression of the orbital angular momentum $\bm{l}$ of Eqs.\eqref{eq:lxmatrixelements}-\eqref{eq:lzmatrixelements} and the Pauli matrices, we can construct the spin-orbit interaction matrix.  Written in the basis
$\Psi^\dagger=\{d_{xz \uparrow}^\dagger,d_{yz \uparrow}^\dagger,d_{xy \downarrow}^\dagger,d_{3z^2-r^2 \downarrow}^\dagger,d_{x^2-y^2 \downarrow}^\dagger,\\
d_{xz \downarrow}^\dagger, d_{yz \downarrow}^\dagger,d_{xy \uparrow}^\dagger,d_{3z^2-r^2 \uparrow}^\dagger,
d_{x^2-y^2 \uparrow}^\dagger \}$ it becomes,
\begin{equation}
\label{eq:HSO}
H_{\rm SOC}=\frac{\zeta}{2}\Psi^\dagger A \Psi,
\end{equation}
where $\Psi^\dagger$ is a row vector, and $\Psi$ is the complex conjugate column vector, and 
\begin{equation}
\resizebox{1\hsize}{!}{$
A=
\begin{pmatrix}
\begin{array}{c|c}
\begin{matrix}
\begin{array}{ccc|cc}
0&-i&i&\sqrt{3}&-1\\
i&0&-1&-i\sqrt{3}&-i\\
-i&-1&0&0&-2i\\
\hline
\sqrt{3}&i\sqrt{3}&0&0&0\\
-1&i&2i&0&0
\end{array}
\end{matrix}& 0 \\
\hline
0  & \begin{matrix}
\begin{array}{ccc|cc}
0&i&i&-\sqrt{3}&1\\
-i&0&1&-i\sqrt{3}&-i\\
-i&1&0&0&2i\\
\hline
-\sqrt{3}&i\sqrt{3}&0&0&0\\
1&i&-2i&0&0
\end{array}
\end{matrix}
\end{array}
\end{pmatrix}$},
\end{equation}
expresses the spin-orbit coupling in the full 10 states of the $t_{2g}$ and $e_g$ manifolds, including spin.  The matrix elements are split into terms that act only on the $t_{2g}$-subspace, $H_{\rm SOC}^{t_{2g}}$, terms that acts only one the $e_g$ subspace, $H_{\rm SOC}^{e_g}$, and terms that have matrix elements between $t_{2g}$ and $e_g$ states, $H_{\rm SOC}^{t_{2g}-e_g}$.
The angular momentum matrix elements in the $e_g$ states are zero. Thus, the matrix elements of the  $H_{\rm SOC}^{e_g}$ are zero as well.

\begin{figure}[t]
    \includegraphics[scale=0.65]{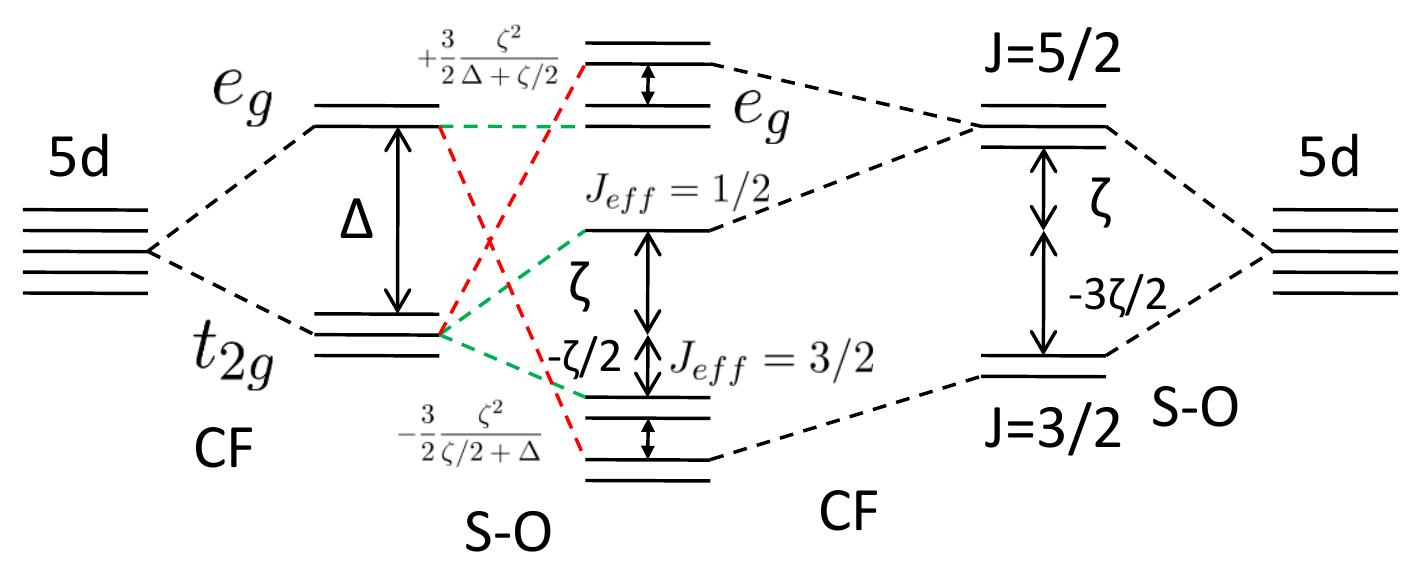}
    \caption{(Color online) Evolution of $d$-orbital states under a cubic crystal field and spin-orbit coupling.  The green lines correspond to the commonly used T-P equivalence that neglects $t_{2g}$-$e_g$ mixing by spin-orbit coupling.  The red lines indicate an extra mixing contribution going beyond TP equivalence, of the $J_{\rm eff}=3/2$ by a factor of $\pm i \sqrt{\frac{3}{2}}\frac{\zeta}{\zeta/2+\Delta} \vert e_g\rangle$, and to the upper quartet $e_g$ by the same factor of $J_{\rm eff}=3/2$ states, as shown in Eqs. \eqref{eq:j3/2modified},\eqref{eq:egmodified}.  The energies of the lower quartet is shifted down by $-\frac{3}{2}\frac{\zeta^2}{\zeta/2+\Delta}$, and of the upper quartet is shifted up by $+\frac{3}{2}\frac{\zeta^2}{\zeta/2+\Delta}$. Notice that the $J_{\rm eff}=1/2$ states are not affected. }
    \label{fig:j_eg}
 \end{figure}

The full Hamiltonian of the one-electron states is
\begin{equation}
 H=H_{\rm SOC}+ H_{\rm CF}.
\end{equation}
In the T-P equivalence one neglects the off-diagonal matrix elements of the angular momentum, $H_{\rm SOC}^{t_{2g}-e_g}$ that connect the $t_{2g}$-$e_g$ subspaces, 
\begin{equation}
\label{eq:TP}
H_{\rm TP}=H_{\rm SOC}^{t_{2g}}+H_{\rm SOC}^{e_g}+H_{\rm CF},
\end{equation}
which is given from the expressions above without the $t_{2g}-e_g$  mixing.  Diagonalizing Eq.\eqref{eq:TP}, the states evolve as shown in Fig.~\ref{fig:j_eg} via the green lines.  In particular, the $e_g$ states are not affected by the spin-orbit coupling, and are separated from the $t_{2g}$ states by an energy difference $\Delta$.  On the other hand, the $t_{2g}$ states are split into eigenstates of energy $\epsilon_{J_{\rm eff}=\frac{1}{2}}=\zeta$:
\begin{equation}
\label{eq:Jeff12}
\begin{split}
\vert J_{\rm eff}= \frac{1}{2},m=-\frac{1}{2} \rangle  &= \frac{1}{\sqrt{3}} \vert d_{yz \uparrow}\rangle -\frac{i}{\sqrt{3}} \vert d_{xz \uparrow}\rangle -\frac{1}{\sqrt{3}} \vert d_{xy \downarrow}\rangle, \\
\vert J_{\rm eff}= \frac{1}{2},m=\frac{1}{2} \rangle  &= \frac{1}{\sqrt{3}} \vert d_{yz \downarrow}\rangle +\frac{i}{\sqrt{3}} \vert d_{xz \downarrow}\rangle +\frac{1}{\sqrt{3}} \vert d_{xy \uparrow}\rangle,
\end{split}
\end{equation}
and eigenstates of energy $\epsilon_{J_{\rm eff}=\frac{3}{2}}=-\frac{\zeta}{2}$:
\begin{equation}
\label{eq:Jeff32}
\begin{split}
\vert J_{\rm eff}= \frac{3}{2},m=-\frac{3}{2} \rangle  &= \frac{1}{\sqrt{2}} \vert d_{yz \downarrow}\rangle -\frac{i}{\sqrt{2}} \vert d_{xz \downarrow}\rangle,\\
\vert J_{\rm eff}= \frac{3}{2},m=\frac{3}{2} \rangle  &= -\frac{1}{\sqrt{2}} \vert d_{yz \uparrow}\rangle -\frac{i}{\sqrt{2}} \vert d_{xz, \uparrow}\rangle,  \\
\vert J_{\rm eff}= \frac{3}{2},m=-\frac{1}{2} \rangle  &= \frac{1}{\sqrt{6}} \vert d_{yz \uparrow}\rangle -\frac{i}{\sqrt{6}} \vert d_{xz \uparrow}\rangle +\sqrt{\frac{2}{3}} \vert d_{xy \downarrow},\rangle \\
\vert J_{\rm eff}= \frac{3}{2},m=\frac{1}{2} \rangle  &=- \frac{1}{\sqrt{6}} \vert d_{yz \downarrow}\rangle -\frac{i}{\sqrt{6}} \vert d_{xz \downarrow}\rangle +\sqrt{\frac{2}{3}} \vert d_{xy \uparrow}\rangle.
\end{split}
\end{equation}
The results in Eq.\eqref{eq:Jeff12} and Eq.\eqref{eq:Jeff32} are commonly used in the literature.  Beyond the T-P equivalence one needs to consider the neglected mixing of the $t_{2g}$-$e_g$ subspaces of the spin-orbit coupling $H_{\rm SOC}^{t_{2g}-e_g}$.  Here, we consider it as a perturbation $H_1=H_{\rm SOC}^{t_{2g}-e_g}$ to the $H_0=H_{\rm TP}$  T-P equivalence terms of Eq. \eqref{eq:TP}. 

Writing $H_0+H_1$ in the diagonal basis of $H_0$, we have in the basis 
$\Phi^\dagger=\{\vert \frac{1}{2},-\frac{1}{2} \rangle,\vert \frac{3}{2},+\frac{3}{2} \rangle, \vert d_{3z^2-r^2}, -\frac{1}{2} \rangle,  \vert \frac{3}{2},-\frac{1}{2} \rangle, \\
\vert d_{x^2-y^2}, -\frac{1}{2} \rangle, \vert \frac{1}{2}, +\frac{1}{2} \rangle,\vert \frac{3}{2},-\frac{3}{2} \rangle, \vert d_{3z^2-r^2}, +\frac{1}{2} \rangle,  \vert \frac{3}{2},+\frac{1}{2} \rangle, \\
\vert d_{x^2-y^2}, +\frac{1}{2} \rangle \}$,
\begin{equation}
\label{eq:HSOdiag}
H_0+H_1=\frac{\zeta}{2}\Phi^\dagger
B
\Phi
\end{equation}
where $\Phi^\dagger$ is a row vector, and $\Phi$ is a complex conjugate column vector, 
\begin{equation}
\resizebox{1\hsize}{!}{$
B=\begin{pmatrix}
\begin{array}{c|c}
\begin{matrix}
\begin{array}{c|cc|cc}
2&0&0&0&0\\
\hline
0&-1&i\sqrt{6}&0&0\\
0&-i\sqrt{6}&\delta&0&0\\
\hline
0&0&0&-1&-i\sqrt{6}\\
0&0&0&i\sqrt{6}&\delta
\end{array}
\end{matrix}& 0 \\
\hline
0  & \begin{matrix}
\begin{array}{c|cc|cc}
2&0&0&0&0\\
\hline
0&-1&-i\sqrt{6}&0&0\\
0&i\sqrt{6}&\delta&0&0\\
\hline
0&0&0&-1&i\sqrt{6}\\
0&0&0&-i\sqrt{6}&\delta
\end{array}
\end{matrix}
\end{array}
\end{pmatrix}$},
\label{eq:H1}
\end{equation}
where $\delta=2\Delta/\zeta$. Note that $H_0$ are the diagonal matrix elements, and $H_1$ are the non-diagonal ones, of the $B$-matrix, Eq.\eqref{eq:H1}.  One sees that there are no matrix elements involving $\vert J_{\rm eff}=\frac{1}{2},m=\pm \frac{1}{2}\rangle$ states. Thus they remain unaffected.  However, the $\vert J_{\rm eff}=\frac{3}{2}\rangle$ and $e_g$ subspaces are mixed. Thus, going beyond the T-P equivalence involves mixing the upper and the lower states as seen in Fig. \ref{fig:j_eg} indicated with red lines.  Hence the evolution of the $t_{2g}$ and $e_g$ states in the presence of spin-orbit coupling is more complex than the commonly used T-P equivalence assumes.

To first order in the wavefunctions, the lower quartet is modified by
\begin{equation}
\label{eq:j3/2modified}
\begin{split}
&\vert J_{\rm eff}=\frac{3}{2},m=\pm \frac{3}{2}\rangle \pm  i\sqrt{\frac{3}{2}}\frac{\zeta}{\zeta/2+\Delta} \vert d_{3z^2-r^2},\mp \frac{1}{2} \rangle,\\
&\vert J_{\rm eff}=\frac{3}{2},m= \pm \frac{1}{2}\rangle \pm  i\sqrt{\frac{3}{2}}\frac{\zeta}{\zeta/2+\Delta} \vert d_{x^2-y^2},\pm \frac{1}{2} \rangle,
\end{split}
\end{equation}
and to second order in energy we find a shift by $-\frac{3}{2}\frac{\zeta^2}{\zeta/2+\Delta}$.  The upper quartet is modified by
\begin{equation}
\label{eq:egmodified}
\begin{split}
&\vert d_{3z^2-r^2},\mp \frac{1}{2} \rangle \pm  i\sqrt{\frac{3}{2}}\frac{\zeta}{\Delta+\zeta/2} \vert J_{\rm eff}=\frac{3}{2},m=\pm \frac{3}{2}\rangle,\\
&\vert d_{x^2-y^2},\pm \frac{1}{2} \rangle \pm  i\sqrt{\frac{3}{2}}\frac{\zeta}{\Delta+\zeta/2} \vert  J_{\rm eff}=\frac{3}{2},m= \pm \frac{1}{2}\rangle,
\end{split}
\end{equation}
with shifts in energies of $+\frac{3}{2}\frac{\zeta^2}{\Delta+\zeta/2}$.  Note for $\zeta=0.5eV,\ \Delta=3eV$ typical values for 5d systems,
the mixing is $\sqrt{\frac{3}{2}}\frac{\zeta}{\zeta/2+\Delta}\approx0.19$, a 20\% effect.

\section{\label{sec:add_e-e_int}Inclusion of Electron-electron interaction}
Having treated the octahedral crystal field $H_{\rm CF}$ in Sec.~\ref{sec:crystal_field} and the  spin-orbit interaction $H_{\rm SOC}$ in Sec.~\ref{sec:T-P_and_beyond}, we are now ready to add the electron-electron interactions, $H_{\rm e-e}$.  We are especially interested in how electron-electron interactions will interplay with the $t_{2g}$-$e_g$ mixing highlighted in the previous section. This mixing is often ignored in the literature.

\subsection{\label{sec:approximations_3d}T-P equivalence in 3d systems}
In the presence of electron-electron interactions, the Hamiltonian of the ion is
\begin{equation}
\label{eq:Hamiltonian}
H= H_{\rm CF}+H_{\rm SOC}+H_{\rm e-e},
\end{equation}
which contains the crystal field part $  H_{\rm CF}$, the spin-orbit part  $ H_{\rm SOC}$, and the interacting part $  H_{\rm e-e}$.  
Within the crystal field approximation several different cases arise: weak, intermediate, and strong crystal field.\cite{fazekas, zeiger-and-pratt}
 The simplest is the weak crystal field case, 
\[\mathcal{E}_{\rm e-e} >>  \mathcal{E}_{\rm CF} > \mathcal{E}_{\rm SOC}, 
\]
where the energy of the interacting part $ \mathcal{E}_{\rm e-e}$ is much larger than the crystal field energy terms $ \mathcal{E}_{\rm CF}$, and the spin-orbit coupling is smaller still.  The intermediate crystal field case is
\[
 \mathcal{E}_{\rm e-e} >  \mathcal{E}_{\rm CF} >  \mathcal{E}_{\rm SOC},
\]
which follows the same order, but the crystal fields are no longer much weaker than the electron-electron interactions.

In 3d systems, the on-site Coulomb interaction is on the order of U= 3-10 eV,  crystal fields are $\Delta$=1.5-2 eV, Hund's coupling is $J_H$=0.8-0.9 eV, and the spin-orbit coupling is in the order of 0.01eV-0.1eV ($\lambda$=0.02 eV for Ti, and $\lambda$= 0.07 eV  for heavier Co).\cite{TMO_book}  Thus, 3d systems fall into the weak and intermediate crystal field regimes.
 
Following the above scheme from the most dominant term to the weakest, we have the  interacting Hamiltonian, which is rotationally invariant with spin independent (Coulomb) interactions.  Thus, the orbital angular momentum $L$ and spin $S$ are conserved quantum numbers and can be used to label the states. The next important term, the crystal field, is not rotationally invariant and mixes different $L$ terms.  Because the energy difference of different $L$ terms is 3-10 eV, and the crystal field is 1.5-2 eV, as a first approximation we neglect the mixing of different $L$ values, and we consider the effect of crystal field splitting within the ground state manifold of the $L$ term, following the conventions of the field.  The smallest term in the hierarchy, the spin-orbit coupling, mixes states of different crystal field levels ($t_{2g}$ and $e_g$ in our case), and terms of different $L$ levels as well, but we neglect those and only include the splitting within the ground state multiplet of crystal field split levels. 

Since the electron-electron interaction is the most dominant term in the above hierarchy and the crystal field mixes states within a given $(L,S)$ term, Hund's first and second rule are valid even in the presence of crystal fields. This means that 3d ions can form high spin structures, where the 4$^{th}$ and 5$^{th}$ electrons  go into the $e_g$ orbitals, as indicated from Hund's first rule of maximal spin. The condition for the low-spin to high-spin transition where the 4$^{th}$ electron prefers to go into the $e_g$ orbitals is approximately $\Delta_{\rm CF} \approx 3 J_H$ (larger $J_H$ favors a high-spin configuration, smaller $J_H$ a low-spin configuration).  Since $\Delta_{\rm CF}$=1.5-2 eV and $J_H$=0.8-0.9eV, this condition is satisfied. However, since crystal fields dominate over the spin-orbit coupling, Hund's third rule ceases to apply. This means that though $L$ and $S$ remain valid quantum numbers, and their values are still given by Hund's first and second rule, the total angular momentum $J$ is no longer a good quantum number.

In the case of strong crystal fields,
\[
 \mathcal{E}_{\rm CF} \geq  \mathcal{E}_{\rm e-e} >   \mathcal{E}_{\rm SOC}, 
\]
the crystal fields are comparable to (or larger than) the electron-electron interaction giving rise to Hund's first and second rule.  Thus, they even mix states belonging to different $(L,S)$ terms.  It is quite usual to find strong crystal fields in 4d and 5d transition metal compounds. On the other hand, there are only rare instances of insulating solids where 3d ions are subject to such strong crystal fields that even Hund's first rule is put out of action.  In next section we will more extensively discuss the case of 4d and 5d systems.

Regardless of the particular energy hierarchy that is relevant, one has
\begin{equation}
[H_{\rm e-e}+H_{\rm CF},\bm{S}^2]=0,\;[H_{\rm e-e}+H_{\rm CF},S^z]=0,
\end{equation}
so that $\bm{S}^2$ and $S^z$ commute with $H_{\rm e-e}$ and $H_{\rm CF}$ since they are spin independent.  As a consequence,  $H_{\rm e-e}+H_{CF}$ has a ground state with well defined spin quantum number.  This holds for arbitrary strength of the Coulomb interaction (including none at all).

Summarizing, the ground state multiplet of $H_{e-e} + H_{\rm CF}$ is only $t_{2g}$ (for up to 6 electrons) if the ion is in the low spin configuration. 
For finite spin-orbit coupling, $S$ and $S^z$ are no longer good quantum numbers.  As discussed in Sec.~\ref{sec:T-P_and_beyond}, $H_{\rm SOC}$ splits  into $H_{\rm SOC}^{t_{2g}}+ H_{\rm SOC}^{t_{2g}-e_g}$ ($H_{\rm SOC}^{e_g}=0$).  Since in 3d systems the spin-orbit coupling is on the order of 0.02-0.07 eV and crystal fields $\Delta =1.5-2$eV, the mixing of $t_{2g}$ and $e_g$ states in the low-spin configuration will be on the order of $\zeta/\Delta \approx 0.02eV/2 eV= 1/100$ and can be neglected to first order.  Consequently, it is a good approximation in 3d systems to neglect the off-diagonal matrix elements of angular momentum in $t_{2g}$ systems and use the T-P equivalence.  This is no longer the case for the heavier transition elements.

\subsection{\label{sec:approximations_4d_5d}Limitations of the T-P equivalence in 4d and 5d systems}

As one moves from 3d to 4d to 5d transition metals the outermost electronic wavefunctions become more and more extended, and thus scale of the typical Hubbard $U$ becomes smaller, reaching down to $U$=0.5-3eV in 5d elements.  The Hund's coupling is reduced as well, to $J_H$=0.6-0.7 eV in 4d elements and to $J_H$=0.5eV in 5d elements. Similarly, the larger spatial extent of the outermost electronic states increase the crystal field splitting to  $\Delta$=1-5eV in 5d elements.  Heavier elements have larger spin-orbit coupling, and its value is increased to $\zeta=0.1-1$eV in 5d elements.  These values bring the 4d/5d elements into the strong crystal field scenario mentioned in the previous section, where the energy scale of the crystal fields is greater than or comparable to the electron interactions. 

Since $\mathcal{E}_{e-e} \approx \mathcal{E}_{\rm CF}$ there is mixing of $(L,S)$ terms. Due to stronger crystal fields and smaller Hund's coupling $J_H$, even Hund's first rule of maximal spin is violated in 4d and 5d systems.  Since $\Delta<3J_H$ (the approximate criterion with $\Delta=3$eV,$J_H$=0.5eV) is not satisfied, a low-spin $t_{2g}$ ground state configurations are preferred. However, a crucial difference of 4d/5d systems relative to their 3d counterparts is the strong spin-orbit coupling.

To help understand the relevant physics, it is useful to briefly consider 4f systems where, 
\[
\mathcal{E}_{\rm e-e} > \mathcal{E}_{\rm SOC}  >  \mathcal{E}_{\rm CF},
 \]
since the spin-orbit coupling is greater than crystal fields, Hund's third rule, takes precedence over lattice effects. Crystal field mixing of different $J$-manifolds are dropped in a first approximation and crystal field effects are considered only within a given $J$-manifold.  

Returning to 5d systems, we have the following hierarchy:
\[
\mathcal{E}_{\rm CF} \approx  \mathcal{E}_{\rm e-e}  \gtrapprox \mathcal{E}_{\rm SOC}.
\]
In this scenario, which occurs mainly in 5d systems and is intermediate to 3d systems and 4f systems, all energy scales are comparable, with spin-orbit coupling  smaller, but still the same order of magnitude as the others. None of the approximations used in 3d and 4f systems work in this regime.  Therefore, in order to study this regime in detail we turn to an exact diagonalization study.

As mentioned in Sec.\ref{sec:T-P_and_beyond}, the off-diagonal elements of spin-orbit coupling mix the $t_{2g}$ and $e_g$ states.  In 5d systems spin-orbit coupling is an order of magnitude greater than 3d systems, and although crystal fields are larger as well, they remain of the same order of magnitude.  Thus, the first order correction in perturbation theory of the wavefunction due to $t_{2g}$-$e_g$ mixing coming from the off-diagonal elements of the spin-orbit coupling is of the order of $\zeta/\Delta \approx 0.5/3= 1/6$. When electron-electron interaction is present, the competition between the Hund's coupling $J_H$, and the crystal field strength $\Delta$ will reduce further the energy difference between low spin states(of $t_{2g}$ only) and high spin states($t_{2g}-e_g$) inducing further mixing. Therefore, it is not as small as in 3d systems and neglecting the $e_g$ states by using the T-P equivalence will result in more dramatic differences from the full $t_{2g}$-$e_g$ space of states.

\section{\label{sec:Model_and_calculations}Model and calculations}
To study the mixing between $t_{2g}$ and $e_g$ orbitals, we use a five-orbital model, taking in account all the $d$-orbitals. Depending on the electron filling, we compare the five-orbital model with a three-orbital $t_{2g}$-only model, or to a two-orbital $e_g$-only model.  We compute various observables as a function of the mixing parameter (of $t_{2g}$ and $e_g$ states), which is the bare spin-orbit coupling strength, $\zeta$. We do this for every electron filling, from one electron to nine electrons.

We model the electron-electron interaction with the Kanamori Hamiltonian,\cite{TMO_book,Kanamori_1963}
\begin{equation}
\label{eq:Kanamori}
\begin{split}
&H^{(Kanamori)}\\
&=U\sum_{m}{\hat{n}_{m \uparrow}\hat{n}_{m \downarrow}+U'\sum_{m\neq m'}{\hat{n}_{m \uparrow}\hat{n}_{m' \downarrow}}}\\
&+(U'-J_H)\sum_{m<m',\sigma}{\hat{n}_{m\sigma}\hat{n}_{m'\sigma}}-J\sum_{m\neq m'}{d_{m\uparrow}^\dagger d_{m\downarrow} d_{m'\downarrow}^\dagger d_{m'\uparrow}}\\
&+J_H\sum_{m\neq m'}{d_{m\uparrow}^\dagger d_{m\downarrow}^\dagger d_{m'\downarrow} d_{m'\uparrow}},
\end{split}
\end{equation}
where $d(d^\dagger)$ is the electron annihilation(creation) operator, $d_{yz},d_{zx},d_{xy},d_{3z^2-r^2},d_{x^2-y^2}$ are associated with labels $m,m'=1,2,3,4,5$ respectively, and  $\hat{n}_{m\sigma}\equiv d_{m\sigma}^\dagger d_{m\sigma}$. For the three orbital $t_{2g}$-only model $m,m'=1,2,3$ and for the two orbital $e_g$-only model $m,m'=4,5$. We assume that the relation $U=U'+2J_H$ is satisfied, which is a good approximation for many materials.\cite{TMO_book}  We take $U'=1$eV in all calculations, leaving only one free parameter, the Hund's coupling $J_H$. For the five-orbital model, Eq.\eqref{eq:Kanamori} is supplemented by 
$H_{\rm CF}$, which is given in the Eq. \eqref{eq:Hcf}. The full Hamiltonian we consider is then 
\begin{equation}
\label{eq:Kan_full}
H=H^{(Kanamori)}+H_{\rm CF}+H_{\rm SOC},
\end{equation}
with $m,m'=$1-5. For the three-orbital $t_{2g}$-only model $H=H^{(Kanamori)}+H_{\rm SOC}^{t_{2g}}$ with $m,m'=1,2,3$,  and for the two-orbital $e_g$-only model $H=H^{(Kanamori)}$ with $m,m'=4,5$. Using exact diagonalization we will compare the results of the full Hamiltonian in Eq.\eqref{eq:Kan_full} with the  $t_{2g}$-only model and the $e_g$-only model.

We calculate expectation values of different operators  $ \hat O, \ O\equiv\langle \psi_0\vert \hat O \vert \psi_0 \rangle$, where $\psi_0$ is the ground state of the many-electron system.  We compute the expectation value of the total spin angular momentum $\bm{S}^2$, the total orbital angular momentum $\bm{L}^2$,  the zero, the single, and the double occupancies of different orbitals defined by\cite{Matsuura}
\begin{eqnarray}
&\hat Z_i \equiv 1-n_{i\uparrow}-n_{i\downarrow}+n_{i\uparrow}n_{i\downarrow},\\
&\hat S_i \equiv n_{i\uparrow}+n_{i\downarrow}-2 n_{i\uparrow}n_{i\downarrow},\\
&\hat D_i \equiv  n_{i\uparrow}n_{i\downarrow},
\end{eqnarray}
where $i$ stands for the orbital index. The amplitudes of the spin, orbital, and total angular magnetic moments, respectively, are defined by $M_s/\mu_B=\vert \sum_i s_z^i \vert$, $M_l/\mu_B=\vert \sum_i l_z^i \vert$, and  $M_{tot}/\mu_B=\vert \sum_i (l_z^i+s_z^i) \vert$, where $s_z^i$ and $l_z^i$ are the z components of the spin and orbital angular momenta of the $i^{th}$-electron respectively, and the effective spin-orbit interaction is
\begin{eqnarray}
\overline{\zeta}=-\frac{1}{\zeta} H_{\rm SOC},\\
 \overline{\zeta}_{t_{2g}}=-\frac{1}{\zeta} H_{\rm SOC}^{t_{2g}} ,\\
  \overline{\zeta}_{t_{2g}-e_g}=-\frac{1}{\zeta} H_{\rm SOC}^{t_{2g}-e_g},
\end{eqnarray}
where $\overline{\zeta}$ is in units of $\hbar^2$.

We note that the effective spin-orbit coupling can be probed experimentally through X-ray absorption spectroscopy (XAS) measurements.\cite{Thole_Laan_PRA, Thole_Laan_PRB, Thole_Laan_PRL}  Core electrons from the occupied states $2p_{1/2}$ and $2p_{3/2}$ are excited to the unoccupied states $5d_{3/2}$ and $5d_{5/2}$, respectively, since these are allowed from the selection rules $\Delta J=0,\pm 1$. These absorption processes are referred to as the intensity peaks $I_{L_2}$ and $I_{L_3}$, respectively. 
Van de Laan and Thole\cite{Thole_Laan_PRA, Thole_Laan_PRB, Thole_Laan_PRL} have shown that the ratio of the integrated intensities (area) of the peaks, $BR=I_{L_3}/I_{L_2}$ [called the branching ratio (BR)] is directly related to the ground state expectation value of the spin-orbit coupling $\langle \bm{L\cdot S}\rangle$ (which we call  $\overline{\zeta}$),  through the relation $BR=(2+r)/(1-r)$, where $r=\langle \bm{L\cdot S}\rangle/\langle n_h\rangle$, and $\langle n_h\rangle$ is the average number of holes in the unoccupied $d$-states (including the full five $d$ orbitals), which is approximately valid even in case of strong crystal fields, and particularly when $\Delta \gg \zeta$ \cite{Thole_Laan_PRB}.
When the spin-orbit coupling is zero, the $J$=3/2 and $J$=5/2 $d$-states are degenerate  (see right side of Fig.~\ref{fig:j_eg}), and the ratio of the intensities $I_{L_3}/I_{L_2}$ is equal to the ratio of the occupied states $2p_{3/2}$ and $2p_{1/2}$ which is 2:1. This yields a branching ratio of $BR=I_{L_3}/I_{L_2}=2$.
A deviation from this value is a clear indication of strong spin-orbit coupling, and can give information on the nature of the ground state.  

Since the effective spin-orbit coupling is a local property of the ion, a single-site calculation is expected to capture the essential physics of the experimental measurements.
In our exact diagonalization (ED) calculations, we place an infinitesimal magnetic field in the z-direction, $H^z$ of the order of $10^{-6}$ eV, in order to lift the degeneracy of the ground state, and obtain a unique expression for the eigenvectors of the ground state.  We have verified this small value does not numerically change the expectation values we compute.

\section{\label{sec:Results}Exact Diagonalization Results}

\subsection{\label{sec:5-orbit_vs_t2g_model}Comparison of $t_{2g}$-$e_g$ model with  $t_{2g}$ only model}
For electron filling from one to six electrons, we will compare the results of the full $t_{2g}$-$e_g$ model with the $t_{2g}$ only model.

\subsubsection{\label{sec:1e}1 electron}

 \begin{figure}[t]
    \includegraphics[scale=0.9]{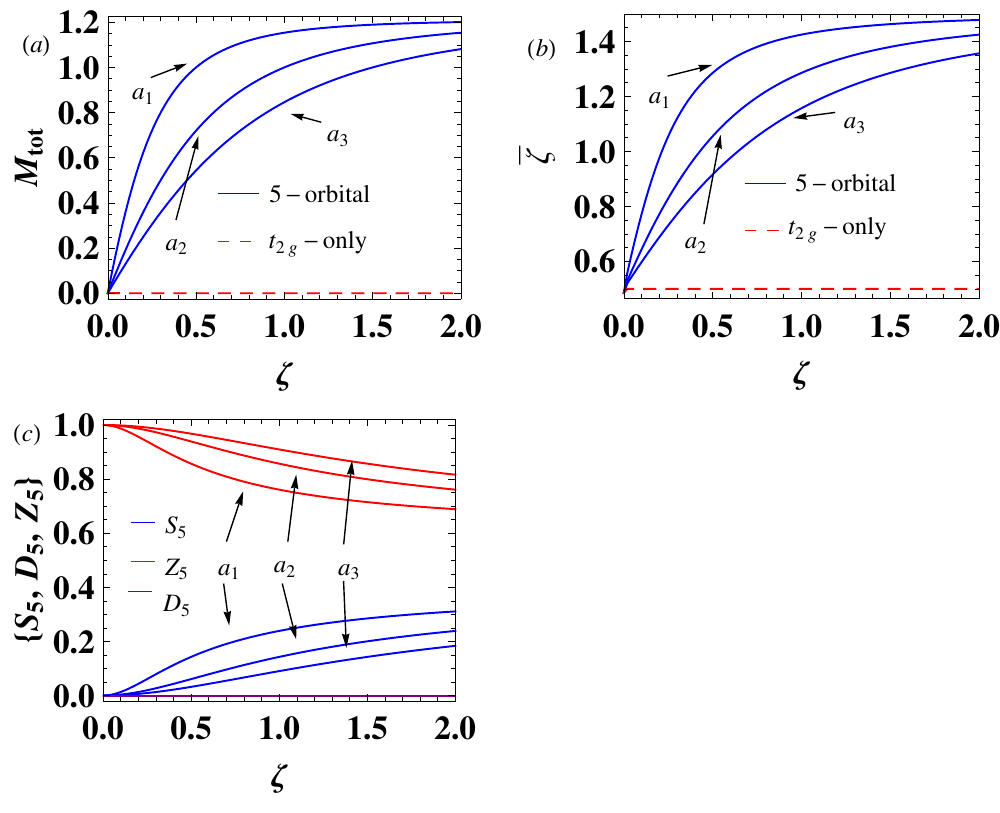}
    \caption{(Color online) Exact diagonalization 1 electron results. (a) Total magnetic moment, $M_{tot}$, (b) effective spin-orbit coupling, $\overline{\zeta}$, (c) single $S_5$, zero $Z_5$, and double $D_5$ occupancies of the $e_g$ $d_{x^2-y^2}$-orbital, for different crystal field values $a_1:\Delta=1$ eV,$a_2:\Delta=2$ eV, and $a_3:\Delta=3$ eV.  Note there is substantial enhancement of the total magnetic moment and effective spin-orbit coupling in the $t_{2g}$-$e_g$ model relative to the $t_{2g}$ only model.}
    \label{fig:fig_1e}
 \end{figure}

In the $t_{2g}$-only model, we have $l=1$ for the orbital angular momentum, and s=1/2. Thus, there is no magnetic moment M=-l+2s=0, since due to spin-orbit coupling, orbital angular momentum and spin angular momentum favor an antiparallel alignment. This is what we see in Fig.\ref{fig:fig_1e}(a).  However,  the quenching of the orbital angular momentum is overestimated in the $t_{2g}$-only model. As we see in the 5-orbital model (for which $l=2$), the restoration of orbital angular momentum due to spin-orbit coupling becomes significant.  
We compute the total magnetic moment for crystal field energy $\Delta=1,2,3$ eV and find it is reduced as the crystal field splitting is increased.  A significant moment remains, for example, for $\Delta=3$ eV and $\zeta=0.5$ eV.  

As shown in Sec.~\ref{sec:T-P_and_beyond} using perturbation theory for a single electron, the off diagonal $t_{2g}$-$e_g$ matrix elements of the spin-orbit coupling creates a small occupancy of $e_g$-orbitals in the ground state. This is seen in Fig.\ref{fig:fig_1e}(b), with the single, zero, and double $e_g$-occupancy of the $e_g$ $x^2-y^2$-orbital, for three different crystal field energies $\Delta=1,2,3$ eV (the single, zero, and double $e_g$-occupancy of the $3z^2-r^2$-orbital are zero). As expected, the occupancies are reduced as the crystal field energy is increased, and they are increased as the spin-orbit coupling strength is increased.
In Fig.\ref{fig:fig_1e}(c) we see for the $t_{2g}$ only model $\overline{\zeta}_{t_{2g}}=0.5$, coming from $\frac{1}{\zeta}\langle H_{SO}^{t_{2g}}\rangle$ in the $\vert J=3/2\rangle$ ground state. In the 5-orbital model, by using Eq.\eqref{eq:j3/2modified} in  calculating the extra contribution from $\frac{1}{\zeta}\langle H_{SO}^{t_{2g}-e_g}\rangle$ of the off-diagonal matrix elements of matrix B in Eq.\eqref{eq:H1}, we get $\frac{1}{\zeta}\langle H_{SO}^{t_{2g}-e_g}\rangle=3\frac{\zeta}{\zeta/2+\Delta}$, thus  $\overline{\zeta}\equiv -\frac{1}{\zeta}\langle H_{SO} \rangle=-\frac{1}{\zeta}\langle H_{SO}^{t_{2g}} \rangle-\frac{1}{\zeta}\langle H_{SO}^{t_{2g}-e_g} \rangle=0.5+3\frac{\zeta}{\zeta/2+\Delta}$ which gives the correct trend shown in Fig.\ref{fig:fig_1e}(c), explaining the missing part not captured from the $t_{2g}$-only model. 

 \begin{figure}[t]
    \includegraphics[scale=0.8]{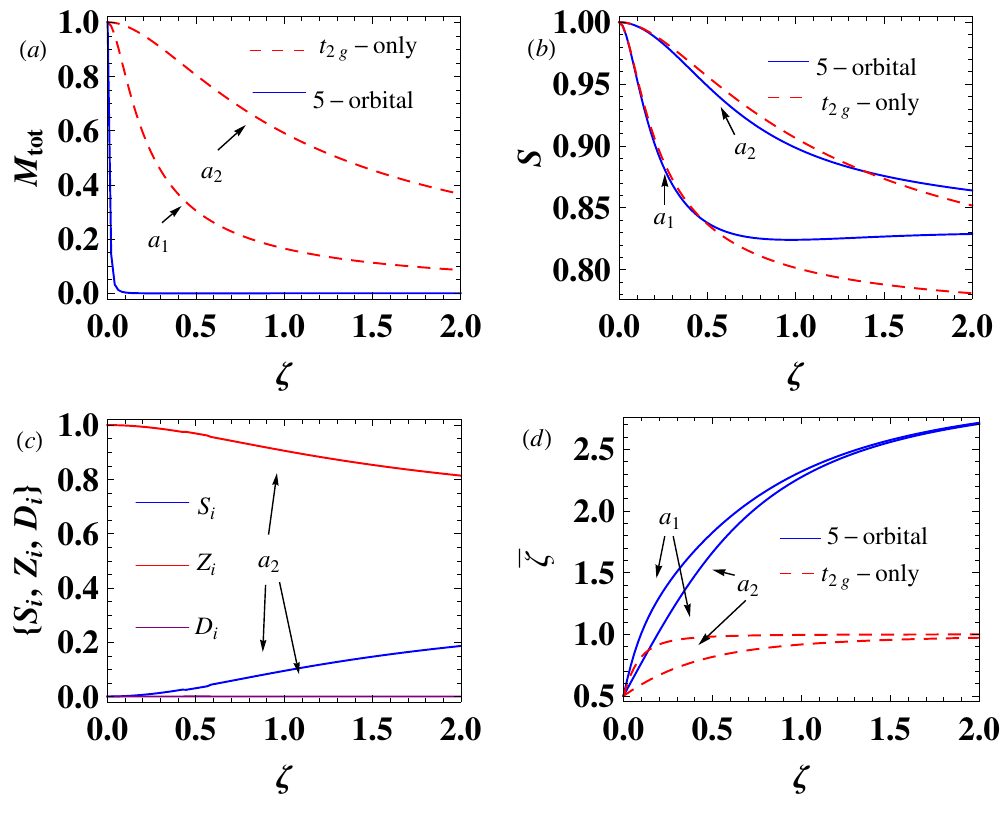}
    \caption{(Color online) Exact diagonalization 2 electron results for crystal field splitting $\Delta=3$ eV. (a) Total magnetic moment $M_{tot}$, (b) spin quantum number $S$, (c) single $S_i$, zero $Z_i$, double $D_i$ occupancy per $e_g$-orbital, (d) effective spin-orbit coupling $\overline{\zeta}$. Different Hund's coupling parameters $a_1: J_H=0.1 eV, a_2: J_H=0.5 eV$ are used.}
    \label{fig:fig_2e}
 \end{figure}

\subsubsection{\label{sec:2e}2 electrons}
In the $t_{2g}$-only model, for zero spin-orbit coupling ($\zeta=0$) $l=1$ and $s=1$. Thus, a non-zero magnetic moment $M_{tot}=-l+2s=1$ is achieved. However, for $\zeta=0$ the 5-orbital model  gives $l=2.7$ because the crystal field mixes different $(L,S)$ terms (with the same $s=1$ as the $t_{2g}$-only model, following Hund's first rule) as discussed in Sec.~\ref{sec:approximations_4d_5d}. At $\zeta=0$  one has the same total magnetic moment as with the $t_{2g}$-only model, $M_{tot}=l_z+2s_z=1$. 

However, when the spin-orbit coupling is turned on, $l_z=0$ and $s_z=0$, so the magnetic moment abruptly plunges to zero, consistent with the approximate rule $l\approx2$, $s=1$, $M_{tot}=-l+2s=0$. In Fig.\ref{fig:fig_2e}(a) we see for the $t_{2g}$-only model with $J_H=0.1$ eV the magnetic moment is reduced as the spin-orbit coupling is increased.   This can be understood as a competition with the Hund's coupling aligning the spins of the electrons, while the spin-orbit coupling ``unaligns" them as it tries to align the spin with the orbital motion. Thus, for  $J_H=0.5$ eV where Hund's coupling is stronger, the effect of the spin-orbit coupling is weaker. 

In Fig.\ref{fig:fig_2e}(b) we see the spin quantum number $S$, for $J_H=0.1,0.5$ eV for the $t_{2g}$-only and for the 5-orbital model as a function of the spin-orbit coupling.  We see that for the smaller Hund's coupling the reduction of the spin is greater, due to the same explanation given for the magnetic moment. The two models match for small spin-orbit coupling, but for $J_H=0.1$ eV a deviation between them appears for $\zeta>0.5$ eV. In Fig.~\ref{fig:fig_2e}(c) we see the single, zero and double $e_g$ occupancy per $e_g$ orbital, for crystal field energy $\Delta=3$ eV and $J_H=0.5$ eV is increased as the spin-orbit coupling is increased. While the curves are similar to the one-electron case, the total result is roughly doubled since it is per $e_g$-orbital.

In Fig.\ref{fig:fig_2e}(d) the effective spin-orbit coupling $\overline\zeta$ is shown for $J_H=0.1,0.5$ eV for the $t_{2g}$-only model and for the 5-orbital model. As the Hund's coupling is increased, the effective spin-orbit coupling is decreased.  As the crystal field is increased, the results from the two models approach each other.   However, $\overline\zeta$ is quite robust even for $\Delta=3$ eV, $\zeta=0.5$ eV, and $J_H=0.1$ eV where the $t_{2g}$-only model gives $\overline{\zeta} \approx 1$ and the 5-orbital model gives $\overline{\zeta} \approx 1.8$.

We can understand these results qualitatively using a single particle analysis. By taking the ground state to be a tensor product of the single-particle eigenstates given in Sec.~\ref{sec:T-P_and_beyond} for the $t_{2g}$-only model and the 5-orbital model, we get for two electrons, 
 $\overline{\zeta}\equiv -\frac{1}{\zeta}\langle H_{SO} \rangle=-\frac{1}{\zeta}\langle H_{SO}^{t_{2g}} \rangle-\frac{1}{\zeta}\langle H_{SO}^{t_{2g}-e_g} \rangle=1+2\times 3\frac{\zeta}{\zeta/2+\Delta}$. The weaker the electronic correlations ({\em i.e.} $J_H=0.1$ eV), the closer one gets to this single electron result. Using this result for the $t_{2g}$-only model gives $\overline{\zeta}_{t_{2g}}=-\frac{1}{\zeta}\langle H_{SO}^{t_{2g}} \rangle=1$ and the 5-orbital model gives an extra contribution $\overline{\zeta}_{t_{2g}-e_g}=-\frac{1}{\zeta}\langle H_{SO}^{t_{2g}-e_g} \rangle=2\times 3\frac{\zeta}{\zeta/2+\Delta}$, which for reasonable values in the 5d elements  ({\em i.e}  $\Delta=3\ eV, \zeta=0.5$ eV), gives for the 5-orbital model  $\overline{\zeta}_{5-orbital}=\overline{\zeta}_{t_{2g}}+\overline{\zeta}_{t_{2g}-e_g}=1+0.96=1.96$ close to what is observed in Fig.\ref{fig:fig_2e}(d). We also see that the two models match at $\zeta<0.1$.  Thus, for 3d systems the T-P equivalence is a good approximation even for the most dramatically different expectation value, the effective spin-orbit coupling.

\subsubsection{\label{sec:3e}3 electrons}

 \begin{figure}[t]
    \includegraphics[scale=0.9]{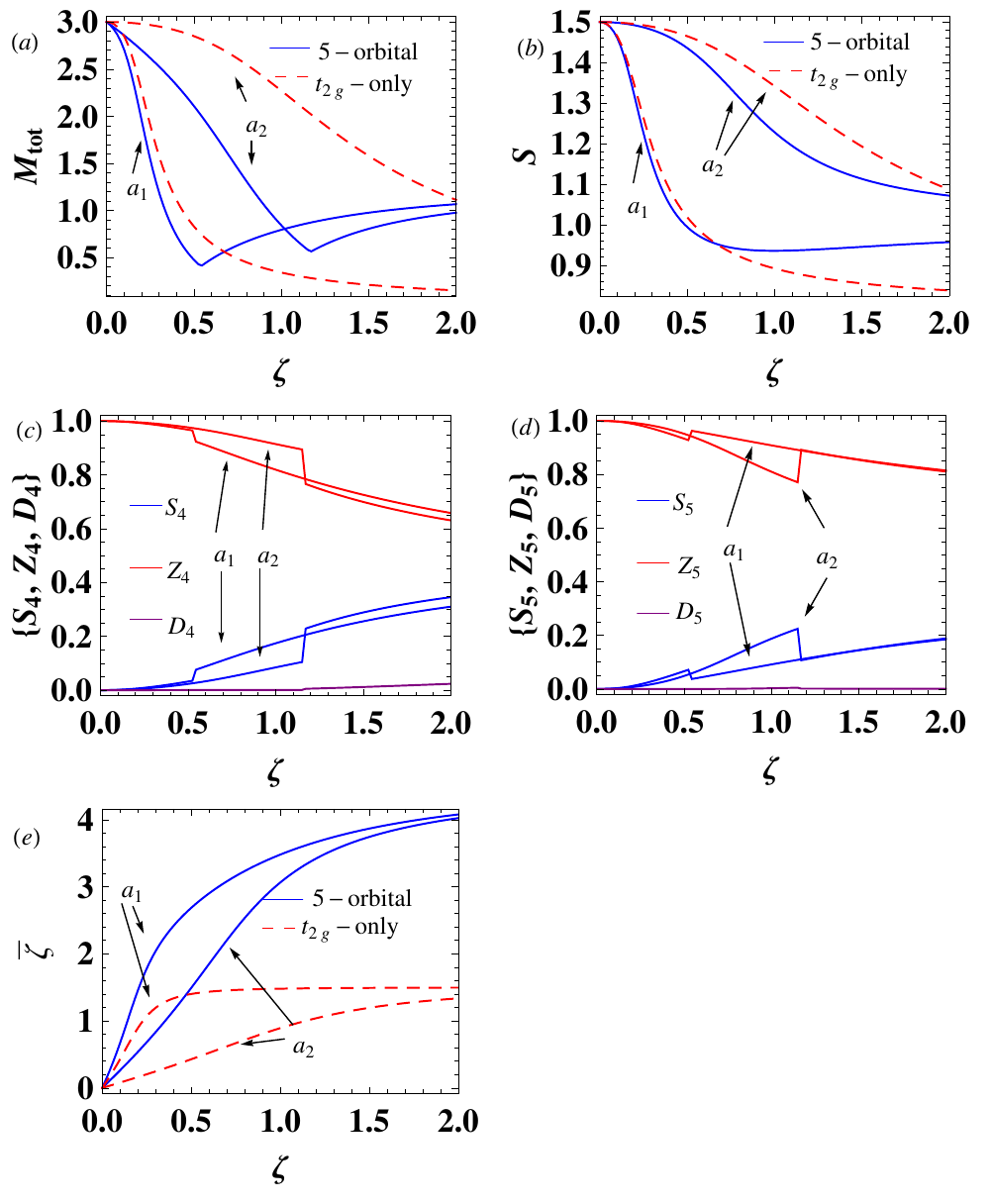}
    \caption{(Color online) Exact diagonalization 3 electron results for crystal field splitting $\Delta=3$ eV. (a) Total magnetic moment $M_{tot}$. (b) Spin quantum number $S$. (c) Single $S_4$, zero $Z_4$, and double $D_4$ occupancies of the $d_{3z^2-r^2}$ orbitals. (d) Single $S_5$, zero $Z_5$, and double $D_5$ occupancies of the  $d_{x^2-y^2}$ orbitals. (e) Effective spin-orbit coupling, $\overline{\zeta}$.  Different Hund's coupling parameters $a_1: J_H=0.1 eV, a_2: J_H=0.5 eV$.}
    \label{fig:fig_3e}
 \end{figure}

For zero spin-orbit coupling for the $t_{2g}$-only model we have $l=0$, and $s=3/2$, while for the 5-orbital model $l=3$ and $s=3/2$, as predicted from Hund's first rule for maximal spin. With this in mind, we turn our attention first to the total magnetic moment, which we expect to reduce with increasing spin-orbit coupling because the spin-orbit coupling tends to ``unalign" the spins. This will be true for both models. However,  comparing our results for the total magnetic moment with Ref.~[\onlinecite{Matsuura}] where a $t_{2g}$-only model was used, we find a significant difference using a 5-orbital model, as seen in Fig.\ref{fig:fig_3e}(a). Thus, the quenching of orbital angular momentum is underestimated in the $t_{2g}$-only model. There is an increased $l_z$ and decreased $s_z$ in the 5-orbital model compared to the $t_{2g}$-only model. When ($\zeta>J_H$) the magnetic moment is reduced rapidly with spin-orbit coupling.  For $J_H=0.1$, when $\zeta$ becomes greater than $J_H$ ($\zeta>J_H$) spin-orbit coupling overcomes the aligning of the spins caused from Hund's coupling.  For $J_H=0.1$ eV there is a transition at $\zeta \approx 0.5$ eV, and for  $J_H=0.5$ eV at $\zeta \approx 1.2$ eV.  The transitions can be seen from the discontinuity in the $e_g$ occupancies where  some small electron occupancy is transferred from one $e_g$ orbital to the other (the average $e_g$-occupancy remains constant). 
There is also some transfer of double occupancy from two $t_{2g}$ orbitals to the third one, where the average $t_{2g}$-occupancy remains constant as well.

As one increases the spin-orbit coupling strength, the total spin is more affected compared to the two-electron system, because it is tightly connected to the orbital angular momentum.  The  $S$ of the $t_{2g}$ and 5-orbital models begin to deviate with increasing strength of the spin-orbit coupling, as seen in the Fig.\ref{fig:fig_6e}(b). For small Hund's coupling this deviation is small, and for larger Hund's coupling this deviation is larger.

For the effective spin-orbit coupling, there is a more dramatic difference between the two models compared to the two-electron system, where for $\zeta=0.5$ eV and $\Delta=3$ eV we have $\overline{\zeta}_{t_{2g}-only}$=1.5 for the $t_{2g}$-only model, while for the 5-orbital model $\overline{\zeta}_{5-orbital}$=2.8.  Using a single particle analysis similar to that of two-electron filling, we get $\overline{\zeta}_{t_{2g}}=1.5,\overline{\zeta}_{5-orbital}=1.5+3\times 3\frac{\zeta}{\zeta/2+\Delta}$, which is very close to what we observe in Fig.\ref{fig:fig_3e}(e) for $J_H=0.1$ eV, while for $J_H=0.5$ eV a significant decrease occurs in the effective spin-orbit coupling.

\subsubsection{\label{sec:4e}4 electrons}

 \begin{figure}[t]
    \includegraphics[scale=0.9]{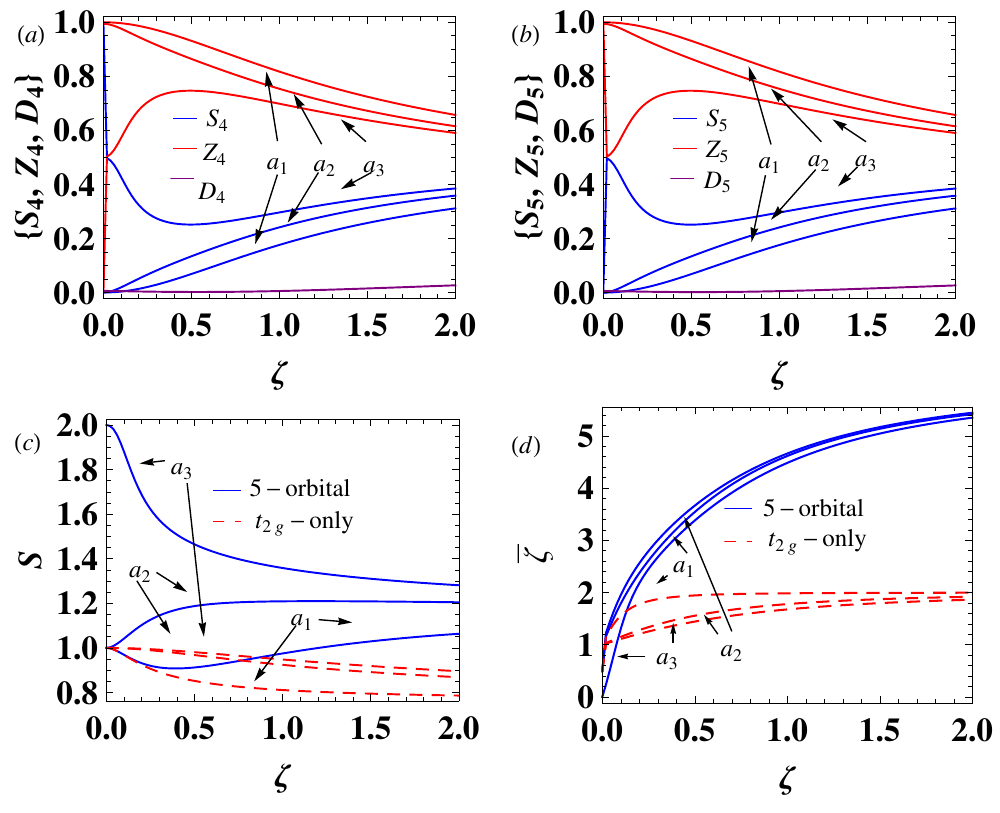}
    \caption{(Color online) Exact diagonalization 4 electron results for crystal field splitting $\Delta=3 eV$. (a) Single $S_4$, zero $Z_4$, and double $D_4$ occupancies of the $d_{3z^2-r^2}$ orbitals.  (b) Single $S_5$, zero $Z_5$, and double $D_5$ occupancies of the $d_{x^2-y^2}$ orbitals. (c)  Spin quantum number, $S$. (d)  Effective spin-orbit coupling, $\overline{\zeta}$. Different Hund's coupling parameters $a_1: J_H=0.1 eV, a_2: J_H=0.5 eV$, $a_3: J_H=0.7 eV$.}
    \label{fig:fig_4e}
 \end{figure}

For four electrons the total magnetic moment is zero in both models: $l_z,s_z=0$. In the $t_{2g}$-only model, $l=1,\ s=1$ and $J=0$ as indicated from the $J=-l+s$ law of the T-P equivalence.
In the five-orbital model there is a low-spin to high-spin transition. For $\Delta=3$ eV at zero spin-orbit coupling and $J_H=0.5$ eV, we find $l=4$, and $s=1$ (low-spin).  While at $J_H=0.7$ eV there is a transition to a high-spin state with $l=2$, and $s=2$.  This can be seen in Fig.\ref{fig:fig_4e} (c) and Fig.\ref{fig:fig_4e}(a). For $J_H=0.7$ eV the fourth electron is shared between the $e_g$-orbitals and the $t_{2g}$-orbitals in a non-monotonic way as a function of spin-orbit coupling. 

In Fig.\ref{fig:fig_4e} (c), for $J_H=0.1$ eV (low-spin) at $\zeta=0$, $s=1$ for both models. However, they start to deviate for $\zeta>0.5$ eV.  For $J_H=0.5$ eV there is a significant deviation between the two models even at small spin-orbit coupling.  At $J_H=0.7$ eV there is a high-spin transition, $s=2$, but there is a rapid reduction of the spin quantum number as a function of spin-orbit coupling, approaching the low-spin value for large $\zeta$.

The effective spin-orbit coupling is seen in Fig.\ref{fig:fig_4e}(d).  We see that the effect of Hund's coupling is weak within each model, although the models show the strong quantitative differences with respect to each other observed at smaller electron numbers. The single electron approach used in smaller electron fillings gives here $\overline{\zeta}_{t_{2g}}$=2, and $\overline{\zeta}_{t_{2g}-e_g}=4\times3\frac{\zeta}{\zeta/2+\Delta}$, giving for the $t_{2g}$-only model $\overline{\zeta}_{t_{2g}-only}$=2, and for $\Delta=3$ eV, and $\zeta=0.5$ eV, giving for the 5-orbital model $\overline{\zeta}_{5-orbital}=\overline{\zeta}_{t_{2g}}+\overline{\zeta}_{t_{2g}-e_g}=2+4\times3\frac{\zeta}{\zeta/2+\Delta}=3.85$, close to what observed in the figure.

\subsubsection{\label{sec:5e}5 electrons}

\begin{figure}[t]
    \includegraphics[scale=0.9]{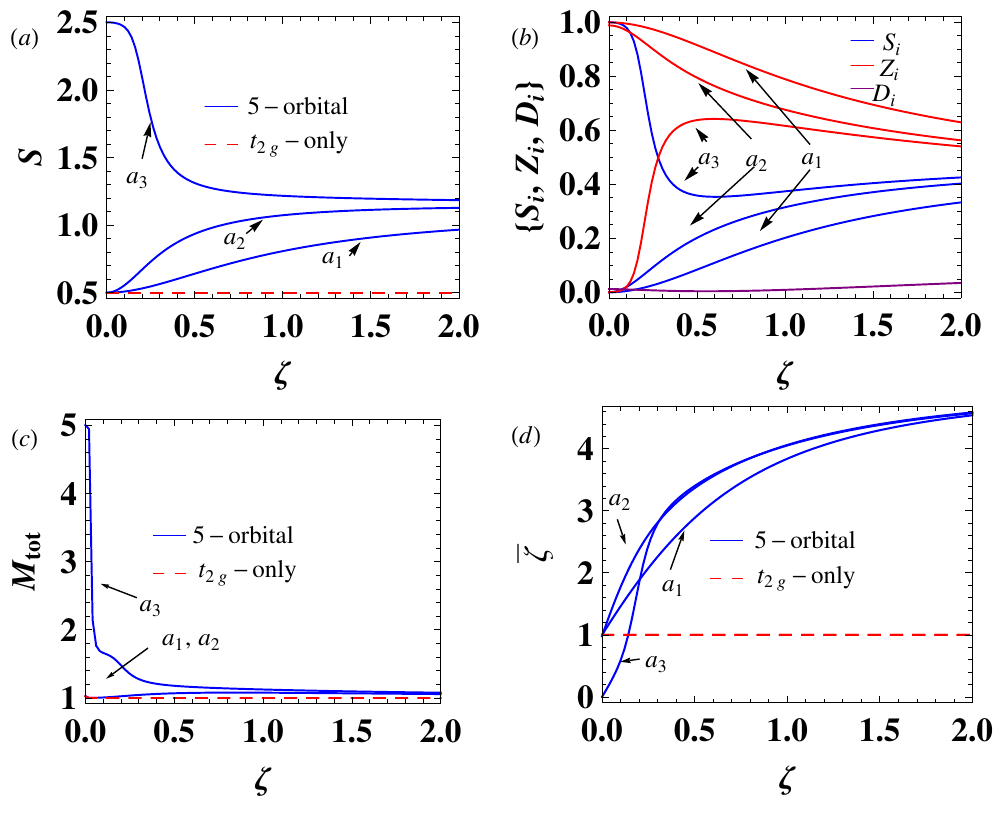}
    \caption{(Color online) Exact diagonalization 5 electron results for crystal field splitting $\Delta=2.7$ eV. (a) Spin quantum number $S$.  (b) Single $S_i$, zero $Z_i$, and double $D_i$ occupancies per $e_g$-orbital. (c) Total magnetic moment $M_{tot}$. (d) Effective spin-orbit coupling $\overline{\zeta}$. Different Hund's coupling parameters $a_1: J_H=0.1 eV, a_2: J_H=0.5 eV$, $a_3: J_H=0.6 eV$.}
    \label{fig:fig_5e}
 \end{figure}

At zero spin-orbit coupling with $\Delta=2.7$ in the five-electron configuration, Fig.\ref{fig:fig_5e}(a)  shows a low-spin configuration $s=1/2$ for $J_H=0.1$ eV and $J_H=0.5$ eV, and a high-spin  $s=5/2$ configuration
 for $J_H=0.6$ eV. Both the high and low-spin configurations evolve continuously as a function of $\zeta$, approaching the same asymptotic value of  $s=1$.
 
The high-spin to low-spin transition is also seen in the $e_g$-occupancies, $S_i,Z_i,D_i$, where $i$ stands for either of the $e_g$-orbitals, plotted in Fig.\ref{fig:fig_5e}(b). For $J_H=0.6$ eV, at zero spin-orbit coupling each $e_g$ orbital is singly occupied. As the spin-orbit coupling is increased, there is a rapid decrease in the $e_g$-occupancies, indicating a high-spin to low-spin transition. However, even in the low-spin case with $J_H=0.5$ eV and $\zeta=0.5$ eV (typical values of 5d systems), there is $S_i=0.2$ single occupancy per $e_g$ orbital, giving a total of 0.4 electrons in the $e_g$-orbitals and an equivalent depletion from the $t_{2g}$-orbitals which cannot be captured from the $t_{2g}$-only model.  

Fig.\ref{fig:fig_5e}(c) shows the total magnetization which stays very close to 1.0, except for the case of  $J_H=0.6$ eV for very small spin-orbit coupling.  The $t_{2g}$-only model gives $M_{tot}=1\mu_B$. In the five-orbital model the low-spin state $J_H=0.5$ eV,  $\Delta=2.7$ eV gives a value very close to that, with slightly reduced $l_z$ and increased $s_z$. The high-spin configuration  $J_H=0.6$ eV, $\Delta=2.7$ eV which at $\zeta=0$ has 5 parallel spins, one in each of the 5-orbitals, starts from $M_{tot}=5\mu_B$, but rapidly reduces to $M_{tot}=1\mu_B$ as the spin-orbit induced high-low spin transition occurs. Thus the state ($J_H=0.6$ eV, $\zeta$=0.5eV, $\Delta=2.7$ eV) which has 0.8 electrons in the $e_g$-orbitals, the state ($J_H=0.5$ eV, $\zeta=0.5$ eV  $\Delta=2.7$ eV) which has 0.4 electrons in the $e_g$-orbitals, and the $t_{2g}$-only state all share the same total magnetic moment $M_{tot}=1\mu_B$. Therefore in this example, the magnetic moment is not a good quantity to distinguish between them.

In Fig.\ref{fig:fig_5e}(d) we see the effective spin-orbit coupling $\overline{\zeta}$. The $t_{2g}$-only model, for which $J_{\rm eff}=1/2$, gives a contribution of $\overline{\zeta}_{t_{2g}}$=1. However in Ref.[\onlinecite{Clancy}], experiments using X-ray absorption spectroscopy in iridium-based compounds in oxygen octahedral fields ($J_H=0.5\ eV,\ \Delta=\ 3eV,\ \zeta=0.5$ eV), a branching ratio BR=6.9 was reported.  This gives an effective spin-orbit coupling $\overline{\zeta}=3.1$,  which is what we find as well within the five-orbital model. 

The authors of Ref.[\onlinecite{Clancy}] emphasize that they find large branching ratios in all Ir compounds studied, with little or no dependence on chemical composition, crystal structure, or electronic state and speculate that unusually strong spin-orbit coupling effects maybe a common feature of all the iridates, or at least those possessing an octahedral local crystal field environment.  These properties are explained well by our model.  First, the effective spin-orbit coupling is a local ion property. Second, an octahedral field environment such as the one studied here shows that the large branching ratio should be a common feature to all the iridates compared.

The authors of Ref.[\onlinecite{Clancy}]  interpret their experimental results as an indication of a $J_{\rm eff}=1/2$ pure state, which has been put forward to explain\cite{Kim1329, Kim_PRL} the insulating properties of Sr$_2$IrO$_4$, and Na$_2$IrO$_3$. In the $J_{\rm eff}=1/2$ scenario, the $J_{\rm eff}=3/2$ band derived from the $J=3/2$ states will be completely occupied, effectively prohibiting any $L_2$ transitions ($2p_{1/2} \rightarrow 5d_{3/2}$) and only $L_3$ transitions will be allowed processes ($2p_{3/2} \rightarrow 5d_{3/2,5/2}$), since the $J_{\rm eff}=1/2$ is separated from the $J=5/2$ states (the lowest unoccupied states). Hence $I_{L_2} \approx 0$, explaining the large branching ratio observed. Whereas in the $S_{\rm eff}=1/2$ scenario, on the other hand, the lowest unoccupied state possesses mixed $J=3/2$ and $J=5/2$ character that allows both $L_2$ and $L_3$ transitions, having lower a BR. (Recall the $BR=I_{L_3}/I_{L_2}$.)  The authors of Ref.[\onlinecite{Clancy}] suggested that the difference between the two BR can distinguish between the two scenarios, and reveal the nature of the ground state. 

However, in the first case the $e_g$ states have been assumed to be infinitely separated from the $t_{2g}$ ones, which gives pure $J_{\rm eff}=3/2$ and $J_{\rm eff}=1/2$ but as we see in Fig.\ref{fig:j_eg} going beyond the T-P equivalence from the strong spin-orbit coupling side, the octahedral crystal field mixes $J=3/2$ and $J=5/2$, which are not mixed at zero octahedral crystal field.

The reported tetragonal distortions of the octahedral oxygen cages mixes $J_{\rm eff}=1/2$ and $J_{\rm eff}=3/2$ and takes one away from the pure $J_{\rm eff}=1/2$ scenario.  We show in this work that even at large crystal fields of $\Delta=3$ eV, the mixing between $t_{2g}$ and $e_g$ manifolds is not negligible.  Accounting for it can explain the remarkably large BR in a more natural, and more general way, for all the Ir-compounds in an octahedral field. Foyevtsova {\em et al.},\cite{Foyevtsova} study Na$_2$IrO$_3$  using DFT calculations with and without spin-orbit  coupling.  To compare the results of their proposed molecular orbital scenario with experiments, they report $\overline{\zeta}=1.91$ by including the $e_g$ orbitals and $\overline{\zeta}=0.73$ by keeping only the $t_{2g}$ in their calculations, supporting a non-pure $J_{\rm eff}=1/2$ state.  Others have reached similar conclusions regarding the admixture of $e_g$ orbitals.\cite{Lovesey,Chapon, Lovesey2}

Measurements of XAS on BaIrO$_3$,\cite{Laguna-Marco} report a BR=4, which gives a $\overline{\zeta}=2.1$--double the canonical value for the $J_{\rm eff}=1/2$ state that gives $\overline{\zeta}=1$--and they  attribute the larger value to the mixing with the $e_g$ states.  Katukuri {\em et al.}\cite{Katukuri_thesis, Katukuri_paper} using quantum chemistry calculations for several iridate oxides report $\overline{\zeta}\approx2$ where they considered hybridization between $e_g$ orbitals and neighboring oxygen ligands, which reduces the value of $\overline{\zeta}$.
In addition, they report that such large deviations from the canonical value of $\overline{\zeta}=1$ of the $t_{2g}$-only model of $J_{\rm eff}=1/2$ cannot be accounted for without the mixing with the $e_g$ states.
In Ref.~[\onlinecite{Haskel}] XAS measurements for Sr$_2$IrO$_4$ report a BR=4.1 which gives $\overline{\zeta}=2.1$ and the deviation from $\overline{\zeta}=1$  is attributed to the mixing of $t_{2g}$ and $e_g$ states.  In Ref.~[\onlinecite{Boseggia}] x-ray resonant magnetic scattering (XRMS) measurements on BaIrO$_3$ gives a BR=5.45, which gives  $\overline{\zeta}=2.67$.

Closing this discussion of the effective spin-orbit coupling in the literature, and coming back to our calculations, a single particle analysis captures well the observed trend, giving $\overline{\zeta}_{5-orbital}=1+4\times 3\frac{\zeta}{\zeta/2+\Delta}=3.03$, for $\zeta=0.5$ eV and $\Delta=2.7$ eV. In Fig.\ref{fig:fig_5e}(d) in the vicinity of $\zeta=0.5$ the effect of the Hund's coupling is to increase the effective spin-orbit coupling. Also, for $J_H=0.6$ eV, $\overline{\zeta}$ starts from zero because in this high-spin configuration $l=0$ and $s=5/2$ for $\zeta=0$.

\subsubsection{\label{sec:6e}6 electrons}

 \begin{figure}[t]
    \includegraphics[scale=0.9]{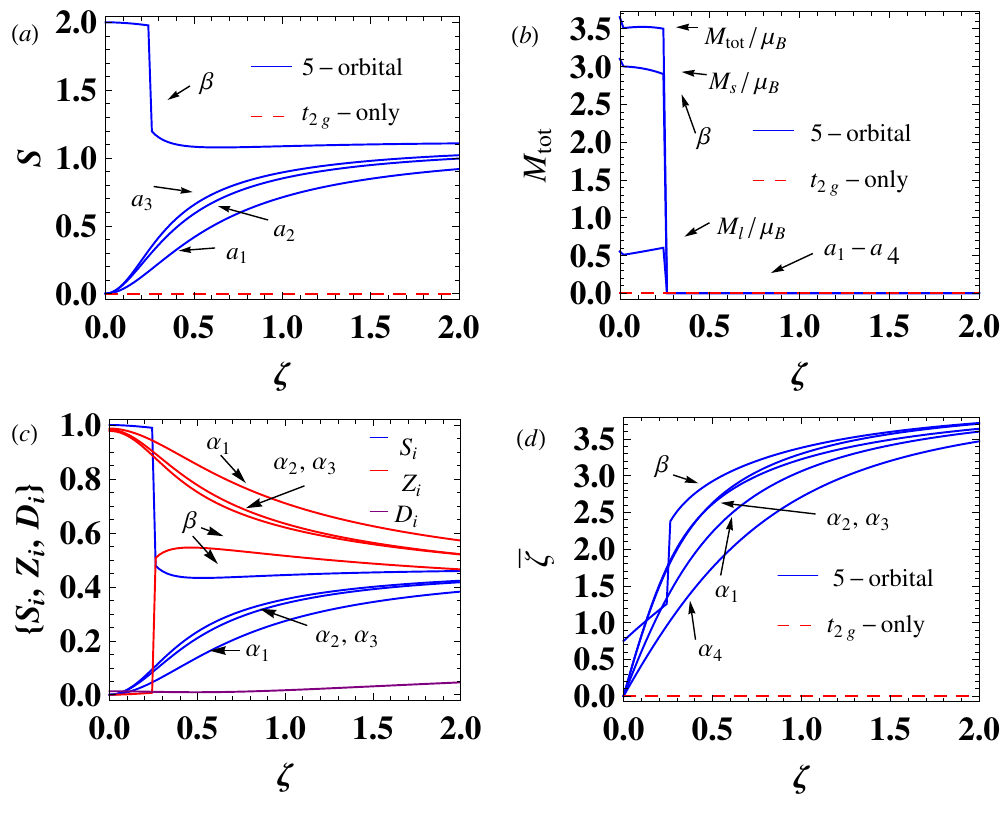}
    \caption{(Color online) Exact diagonalization 6 electron results.  (a) Spin quantum number $S$.  (b) Total magnetic moment $M_{tot})$. (c)  Single $S_i$,  zero $Z_i$, and double $D_i$ occupancy per $e_g$ orbital. (d) Effective spin-orbit coupling $\overline{\zeta}$. Parameters for predominately low-spin configurations: $a_1:\Delta=3 eV, J_H=0.5 eV,a_2:\Delta=2.5 eV, J_H=0.5 eV,a_3:\Delta=3 eV, J_H=0.7 eV,a_4:\Delta=3 eV, J_H=0.1 eV$. Parameters for predominately high spin-configurations $\beta: \Delta=2.5 eV, J_H=0.7 eV$.}
    \label{fig:fig_6e}
 \end{figure}
 
The six-electron results are shown in Fig.\ref{fig:fig_6e}. For the $t_{2g}$-only model the results are trivial: The spin, total magnetic moment, effective spin-orbital coupling are all zero, since we have 6 electrons completely occupying all the $t_{2g}$ orbitals. However, adding two more orbitals changes the picture.  As we see from Fig.\ref{fig:fig_6e} (a), the spin quantum number at zero spin-orbit coupling is $S=0$, but for finite spin-orbit coupling it deviates from that, reaching $S\approx0.5$ around $\zeta=0.5eV$ for the configurations that have $S=0$ at $\zeta=0$.  The low-spin configurations have completely filled $t_{2g}$ orbitals at $\zeta=0$. These configurations are  $a_1:\Delta=3 eV, J_H=0.5 eV, a_2:\Delta=2.5 eV, J_H=0.5 eV, a_3:\Delta=3 eV, J_H=0.7 eV, a_4:\Delta=3 eV, J_H=0.1 eV$. 

Comparing these cases, one sees that when the ratio $J_H/\Delta$ is increased the spin quantum number increases with increasing spin-orbit coupling. If we  continue increasing this ratio to the configuration $\beta: \Delta=2.5 eV, J_H=0.7 eV$, the system will transition to a high-spin state at zero spin-orbit coupling. However, for the high-spin configuration $\beta$, at $\zeta=0.25 eV$ spin-orbit coupling creates a high-spin to intermediate-spin transition, going from $S=2$ to approximately $S=1$.

Turning our attention now to Fig.\ref{fig:fig_6e}(b), we see that only the high-spin $\beta$ configuration has a net magnetic moment, while all other configurations give a zero total magnetic moment. The total magnetic moment of the $\beta$ high-spin configuration is $M_{tot}/\mu_B$=3.5, where $M_{S}/\mu_B$=3 and $M_{l}/\mu_B$=0.5. But at $\zeta=0.25 eV$ where the spin-orbit coupling induces the high-spin to intermediate-spin transition, the magnetic moment vanishes.  The transition is also reflected in the single and zero occupancies per $e_g$-orbital, shown in Fig.\ref{fig:fig_6e}(c). For the $\beta$ configuration and $\zeta<0.25$ there are 2 electrons, 1 per $e_g$-orbital, while for $\zeta>0.25$ there is 1 electron, 1/2 per $e_g$-orbital. Also, for the low-spin configurations $a_1$-$a_3$ there are 0.4 electrons in the $e_g$ orbitals, 0.2 to each orbital.

The effective spin-orbit coupling is shown in Fig.~\ref{fig:fig_6e} (d). The effect of the Hund's coupling is to increase $\bar \zeta$ in the intermediate spin-orbit coupling region. The spin-orbit induced transition from high-spin to intermediate-spin of the $\beta$ configuration, by a jump at $\zeta=0.25 eV$, doubles its value from $\overline{\zeta}=1.2$ to $\overline{\zeta}=2.4$. The single particle perturbative description gives $\overline{\zeta}=4\times3 \frac{\zeta}{\zeta/2+\Delta}$, and as it is expected to work well at small correlation, it is compared to $J_H=0.1 eV$, and for $\zeta=0.5 eV$ and $\Delta=3 eV$ gives a value of $\overline{\zeta}=1.83$, where the exact result gives $\overline{\zeta}=1.81$.

\subsection{\label{sec:5-orbital_vs_eg model}Comparison of $t_{2g}$-$e_g$ model with $e_g$ only model}

For filling from seven to nine electrons, we will compare the results of the full $t_{2g}$-$e_g$ model with $e_g$-only model. The matrix elements of orbital angular momentum are completely quenched in the $e_g$-only model, and thus the spin-orbit coupling as well.

\subsubsection{\label{sec:7e}7 electrons}

\begin{figure}[t]
    \includegraphics[scale=0.85]{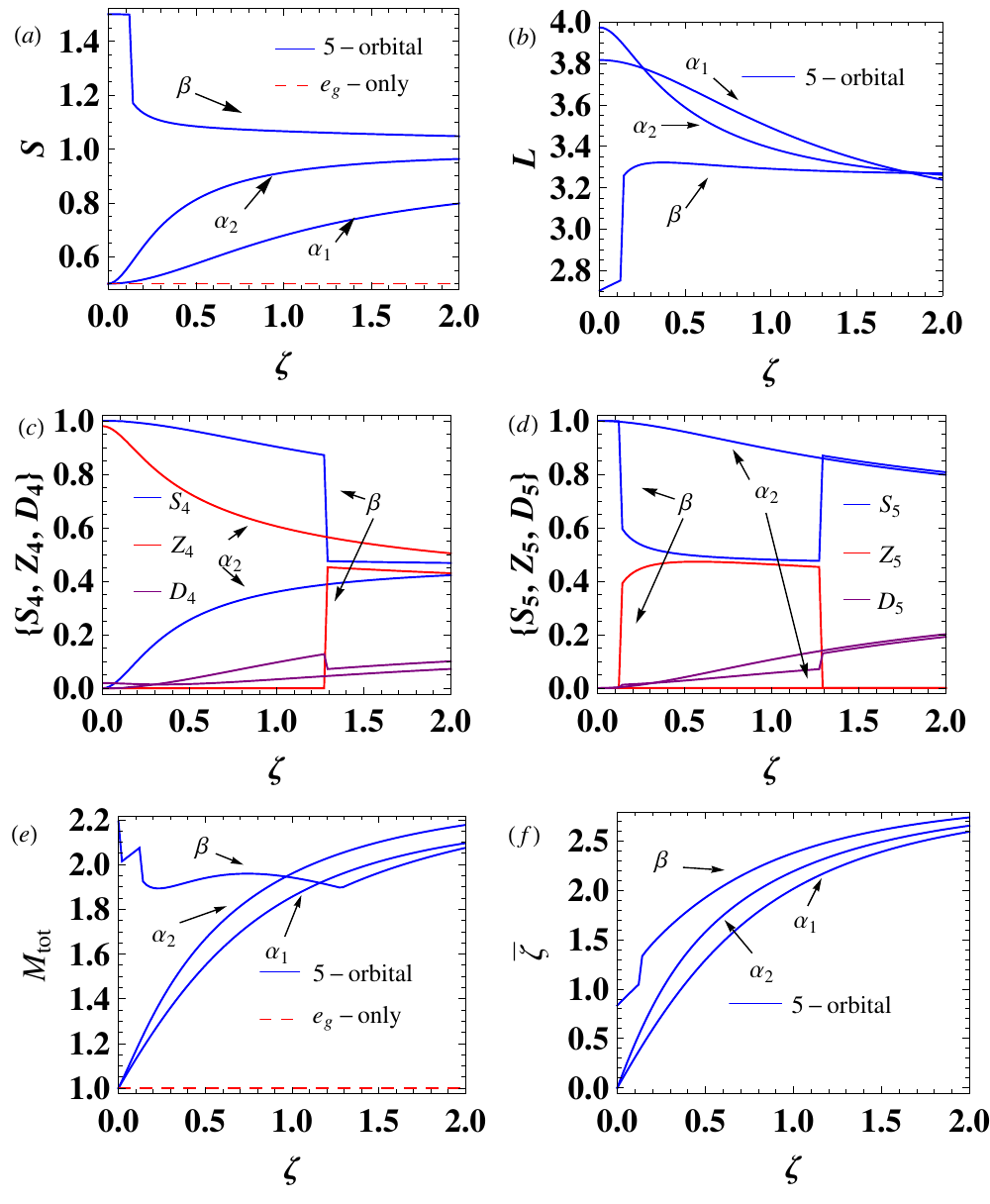}
    \caption{(Color online) Exact diagonalization 7 electron results. (a) Spin quantum number $S$. (b)  Angular momentum quantum number $L$. (c) Single, double, zero occupancies of the $d_{3z^2-r^2}$ orbital ($S_4,D_4,Z_4$). (d) Single, double, zero occupancies of the $d_{x^2-y^2}$ orbital ($S_5,D_5$). (e) Total magnetic moment $M_{tot}$. (f) Effective spin-orbit coupling $\overline{\zeta}$ for $\alpha_1:  \Delta=3 eV, J_H=0.1 eV$, $\alpha_2:  \Delta=2.5 eV, J_H=0.5 eV$, $\beta: \Delta=2.5 eV, J_H=0.7 eV$ configurations. }
    \label{fig:fig_7e}
 \end{figure}

For the seven-electron configuration, we have for the $e_g$-only model a single electron in the $e_g$-orbital, which gives $S=1/2$ as seen in Fig.\ref{fig:fig_7e}(a).  At zero spin-orbit coupling for the configurations $\alpha_1: \Delta=2.5 eV, J_H=0.5 eV$ and $\alpha_2:  \Delta=2.5 eV, J_H=0.5 eV$, $S=1/2$ there is a single electron in the $d_{3z^2-r^2}$ orbital and the rest completely occupy the $t_{2g}$ orbitals, as seen from Fig.\ref{fig:fig_7e}(c), (d). As a function of the spin-orbit coupling, there is a depletion of the $t_{2g}$ orbitals, and an increase in the single occupancy of the $d_{x^2-y^2}$ orbital as seen in Fig.\ref{fig:fig_7e}(d).  This causes an analogous increase in the spin quantum number, as seen in Fig.\ref{fig:fig_7e} (a). When one increases Hund's coupling at zero spin-orbit coupling, there is a low-spin to high-spin transition. In Fig. \ref{fig:fig_7e}(a)  the configurations $\alpha_1:\Delta=3 eV, J_H=0.1 eV$ and $\alpha_2:\Delta=2.5 eV, J_H=0.5 eV$ give $S=1/2$. When the Hund's coupling is increased in the configuration $\beta: \Delta=2.5 eV, J_H=0.5 eV$, we get $S=3/2$ giving two electrons in the $e_g$-orbitals and leaving one hole in the $t_{2g}$ orbitals.  This is shown in Fig.\ref{fig:fig_7e} (c), (d) for the $e_g$-occupancies. At spin-orbit coupling $\zeta=0.12$ the high-spin $\beta$ configuration undergoes an  intermediate-spin transition from $S=3/2$ to $S\approx1.1$ and a subsequent depletion of the $d_{x^2-y^2}$ orbital from 1 electron to 0.5 electron, giving a total 1.5 electrons in the $e_g$-orbitals. At spin-orbit coupling $\zeta=1.3$ eV there is a second transition, interchanging the occupancies between the two $e_g$ orbitals, while keeping the total occupancy of 1.5 electrons in the $e_g$ orbitals constant.  In Fig.\ref{fig:fig_7e}(b) we see the total angular momentum in $\alpha_1, \alpha_2, \beta$ configurations capturing these transitions as well.

\begin{figure}[t]
    \includegraphics[scale=0.87]{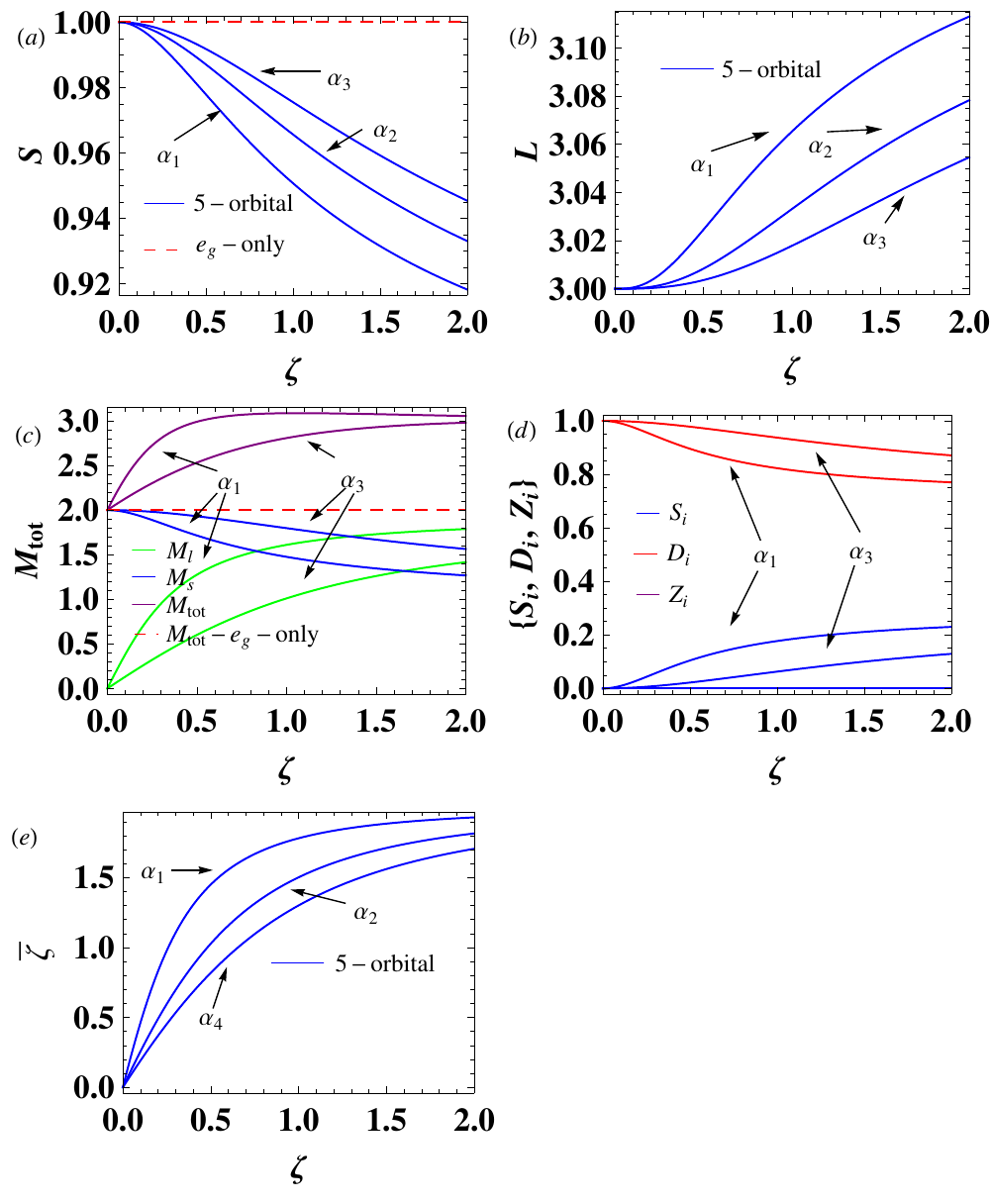}
    \caption{(Color online) Exact diagonalization 8 electron results. (a) Spin quantum number $S$. (b) Angular momentum quantum number $L$. (c) Total magnetic moment $M_{tot}$, orbital magnetic moment $M_l$, and spin magnetic moment $M_S$. (d)  Single $S_i$, double $D_i$, and zero occupancies $Z_i$ per $t_{2g}$-orbital. (e)  Effective spin-orbit coupling ($\overline{\zeta}$) for $\alpha_1:  \Delta=1 eV, J_H=0.5 eV$, $\alpha_2:  \Delta=2 eV, J_H=0.5 eV$, $\alpha_3: \Delta=3 eV, J_H=0.5 eV$, $\alpha_4: \Delta=3 eV, J_H=0.1 eV$ configurations. }
    \label{fig:fig_8e}
 \end{figure}

In Fig.\ref{fig:fig_7e}(e)  the total magnetic moment is shown. For the configurations $\alpha_1, \alpha_2$ there is a significant deviation from the $e_g$-only model in which the orbital angular momentum is completely quenched.  The total moment is only spin.  In the five-orbital model $M_{tot}\approx1.5 \mu_B$ for $\zeta=0.5$, with the difference coming from the orbital magnetic moment $M_l$, since the spin magnetic moment has small deviation from $M_S\approx1 \mu_B$ as a function of spin-orbit coupling.  For the $\beta$ configuration there are two transitions as a function of spin-orbit coupling, which are seen as discontinuities in the $M_{tot}$ Fig.\ref{fig:fig_7e}(e).
 
The effective spin-orbit coupling is shown in Fig.\ref{fig:fig_7e}(f) for three characteristic cases of the low-spin configurations $\alpha_1 (\Delta=3 eV, J_H=0.1 eV)$, $\alpha_2 (\Delta=2.5 eV, J_H=0.5 eV)$ spin, and the high-spin $\beta (\Delta=2.5 eV, J_H=0.7 eV)$ configuration. The single-electron perturbation result gives $\overline{\zeta}=\frac{3}{4}\frac{\zeta^2}{(\zeta/2+\Delta)^2}+3\times3\frac{\zeta}{\zeta/2+\Delta}$ which is close to what is observed in the $\alpha_1$ configuration. Note that the $e_g$-only model gives $\overline{\zeta}=0$, so in 4d and 5d systems with a $d^7$ configuration, a finite effective spin-orbit coupling can be measured.

\subsubsection{\label{sec:8e}8 electrons}

For eight electrons, we naively expect two electrons in the $e_g$ orbitals and the rest are in the completely filled $t_{2g}$ shell.  In Fig.\ref{fig:fig_8e}(a),(b) we see the spin $S$, and orbital angular momentum $L$ quantum numbers, for three different values of the crystal fields, $\alpha_1: \Delta=1 eV, \alpha_2: \Delta=2 eV, \alpha_3: \Delta=3 eV$, all at $J_H=0.5 eV$. The deviation from $S=1$, and $L=3$ is small as a function of spin-orbit coupling. In Fig.\ref{fig:fig_8e}(c) the total magnetic moment $M_{tot}$, the orbital magnetic moment $M_l$, and the spin magnetic moment $M_S$ are plotted, for $\alpha_1$  and $\alpha_3$ configurations. At zero spin-orbit coupling, the orbital angular momentum is completely quenched, as  predicted from the $e_g$-only model. However, spin-orbit coupling gives rise to a significant amount of orbital angular momentum; the smaller the crystal field ($\alpha_1$), the greater the restoration compared to the larger crystal field configuration $\alpha_3$. Spin-orbit coupling causes a small reduction in the spin magnetic moment, and as a result the difference in the total magnetic moment between the five-orbital model and the $e_g$-only model is mainly from the orbital magnetic moment $M_l$. In Fig.\ref{fig:fig_8e}(d) the single $S_i$, double $D_i$, and zero $Z_i$ occupancies per $t_{2g}$-orbital are plotted. The main effect is that there is depletion of the $t_{2g}$ orbitals as a function of the spin-orbit coupling, with a greater effect for smaller crystal fields.

In Fig.\ref{fig:fig_8e}(e), the effective spin-orbit coupling is plotted.  The smaller the crystal field, the less the quenching of the orbital angular moment. Consequently, the effective spin-orbit coupling is larger. The single particle perturbative description gives $\overline{\zeta}=\frac{3}{2}\frac{\zeta^2}{(\zeta/2+\Delta)^2}+2\times3\frac{\zeta}{\zeta/2+\Delta}$, which compared to the least interacting $\alpha_4:\Delta=3 eV, J_H=0.1 eV$ configuration  gives a good qualitative description, of $\overline{\zeta}\approx0.9$ for $\Delta=3$ eV and $\zeta=0.5$ eV.

\subsubsection{\label{sec:9e}9 electrons}

\begin{figure}[t]
    \includegraphics[scale=0.9]{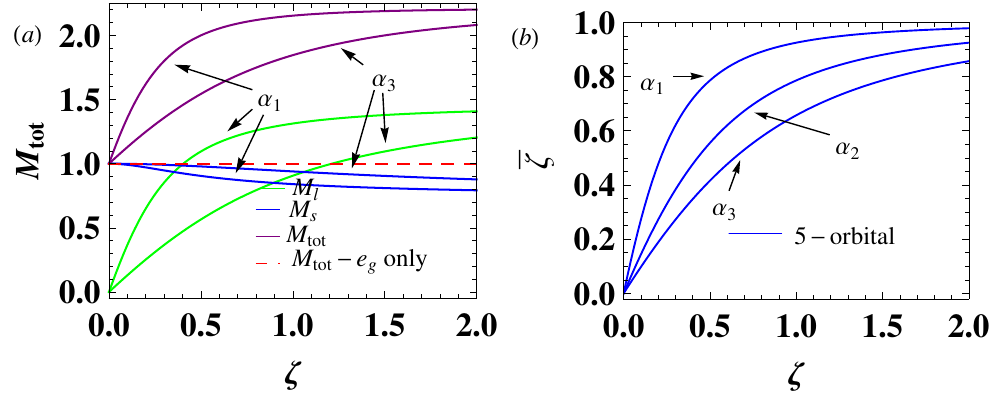}
    \caption{(Color online) Exact diagonalization 9 electron results. (a) Total magnetic moment $M_{tot}$, orbital magnetic moment $M_l$, and spin magnetic moment $M_S$. (b) Effective spin-orbit coupling ($\overline{\zeta}$) for $\alpha_1:  \Delta=1 eV$, $\alpha_2:  \Delta=2 eV$, $\alpha_3: \Delta=3 eV$ configurations. }
    \label{fig:fig_9e}
 \end{figure} 

For the case of nine electrons one has  $S=1/2$ and $L=2$. The angular momentum at $\zeta=0$ is completely quenched giving a total magnetic moment $M_{tot}=1\mu_B$. Spin-orbit coupling gives rise to finite orbital angular momentum. For $\Delta=1$ eV and $\zeta=0.5$ eV one has an extra contribution $M_l=1\mu_B$, and at $\Delta=3$ eV and $\zeta=0.5$ eV one has an extra contribution of $M_l=0.5\mu_B$, as seen in Fig.\ref{fig:fig_9e}(a). The spin magnetic moment $M_S$ is only weakly affected by spin-orbit coupling and remains very close to $M_S=1\mu_B$. The effective spin-orbit coupling is shown in Fig.\ref{fig:fig_9e}(b) for different values of crystal field, $\alpha_1:\Delta=1eV,\alpha_1:\Delta=2eV,\alpha_3:\Delta=3eV$. As the crystal field strength increases, the orbital angular momentum and the effective spin-orbit coupling $\overline{\zeta}$ decreases. The single electron perturbation result gives $\overline{\zeta}=3\times\frac{3}{4}\frac{\zeta^2}{(\zeta/2+\Delta)^2}+3\frac{\zeta}{\zeta/2+\Delta}$, giving for $\Delta=3 eV, \zeta=0.5 eV$ $\overline{\zeta}=0.51$, capturing what we see in Fig. \ref{fig:fig_9e} (b) in $\alpha_3:\Delta=3eV$.  Also there is some small depletion of $t_{2g}$-occupancy due to the $t_{2g}-e_g$ mixing of the off diagonal elements of the spin-orbit coupling interaction.

\section{\label{sec:Summary2}Summary and Conclusions }

In summary, we have carried out an exact diagonalization study of interacting $d$-orbital electrons in a cubic crystal field environment for all electron fillings.  We have focused on mixing effects of the $t_{2g}$ and $e_g$ orbitals induced by the spin-orbit coupling and compared our results to the $t_{2g}$-only and $e_g$-only models commonly used in the literature.  For realistic interaction parameters in Eq.\eqref{eq:Kanamori}, crystal field splitting and spin-orbit coupling Eq.\eqref{eq:HSOdiag}, we find the mixing effects can be significant.  These mixing effects can be important in the interpretation of the branching ratio measured in spectroscopic measurements, which is often used to determine the effective strength of the spin-orbit coupling.  If one assumes a $t_{2g}$-only model (neglecting $t_{2g}$ and $e_g$ mixing) for iridates, for example, one would infer an effective spin-orbit coupling value smaller than the one for the full  $t_{2g}$-$e_g$ model.

For the various electron fillings we calculated the spin $S$, orbital angular momentum $L$, total magnetic moment $M_{\rm tot}$, the single $S_i$, zero $Z_i$, and double $D_i$ occupancy of the $i^{th}$ orbital, and the effective spin-orbit coupling strength $\bar \zeta$.  In general, these quantities can show a complex evolution with the strength of the crystal field splitting $\Delta$ and the bare spin-orbit coupling strength $\zeta$.  For certain electron fillings, crystal field splittings $\Delta$ and Hund's coupling $J_H$, we observe high-spin to low-spin transitions as a function of $\zeta$.  An intermediate spin state may also be realized. The most important results are summarized in Figs.\ref{fig:fig_1e}-\ref{fig:fig_9e}.

In present work we focused on the general effect of octahedral crystal field, however in general the local symmetry is usually lower than the cubic one, and tetragonal or trigonal distortion (of the oxygen cage or due to next neighboring ions) introduces additional complications, which is beyond the scope of the present work. 

The results we have obtained here should be useful in helping to derive more realistic models of local moment interactions in the 4d and 5d transition metal oxides.  These local moment models could then be used to predict what type of magnetic phases and magnetic excitations might be expected in the heavy transition metal oxides.
In this direction we can say that the off-diagonal elements of spin-orbit coupling $H_{SO}^{t_{2g}-e_g}$ can be accounted perturbatively, rather than completely neglected as has been mostly done so far. In that case, we speculate that this effect will probably change the low-energy effective spin models, derived with the use of the T-P equivalence not only quantitatively, but also changes the Hamiltonian structure of each model as well.
Our local moment results could also be used as a starting point for non-equilibrium (Floquet) studies as well since they include an enlarged Hilbert space and can better capture the response of a periodic drive.  These are directions for future research.

\acknowledgements   We gratefully acknowledge funding from ARO grant W911NF-14-1-0579, NSF DMR-1507621, and NSF MRSEC DMR-1720595. This work was performed in part at Aspen Center for Physics, which is supported by National Science Foundation grant PHY-1607611. We would also like to thank  Panteleimon Lapas for fruitful discussions.



%

\end{document}